\documentclass[letter,oneside,9pt,twocolumn,article]{aiaa}
\usepackage{graphicx}
\usepackage{rotating}
\usepackage{bm}
\usepackage{flushend}
\usepackage{cite}
\usepackage[bf,normalsize]{subfigure}
\usepackage{adjustbox}
\usepackage{tabularx}
\usepackage{multirow}
\usepackage{multicol}
\usepackage{indentfirst} 
\usepackage[hidelinks]{hyperref}
\usepackage{float}
\usepackage{xcolor}
\usepackage[round]{natbib}
\setcitestyle{authoryear,open={(},close={)}} 

\usepackage{mwe}
\usepackage[markcase=noupper]{scrlayer-scrpage}
\usepackage{caption} 

\ihead*{{\it Physics of Fluids 38, 036114 (2026), \href{https://doi.org/10.1063/5.0318297}{10.1063/5.0318297}}}

\ohead*{\pagemark}
\ifoot*{\rm }
\ofoot*{\rm Unrestricted content}
\newcommand{\alb}{\vspace{0.1cm}\\} 
\newcommand{\mfd}{\displaystyle}
\renewcommand{\vec}[1]{\bm{#1}}

\renewcommand{\fontsizetable}{\footnotesize}

\renewcommand{\vec}[1]{\bm{#1}}
\newcommand{\Vol}{\rotatebox[origin=c]{180}{\ensuremath{A}}}

\author{Felipe Mart\'in Rodr\'iguez Fuentes\thanks{Graduate Student}~~and~
Bernard Parent\thanks{Associate Professor, bparent@arizona.edu.}\\[0.3em] \it University of Arizona, Tucson, AZ 85721, USA.
}

\title{
Electron Density Depletion in Reentry Plasma Flows\\ Using Pulsed Electric Fields
}

\abstract{ 
Communication blackout due to the plasma layer creates a critical telemetry gap for re-entry vehicles. To mitigate this, we present the first fully-coupled simulation of high-voltage pulsed discharges interacting with a Mach~24 flowfield {using an advanced numerical framework}. The results demonstrate that the applied electric field generates a large, non-neutral plasma sheath near the cathode, depleting electron density by several orders of magnitude over a distance commensurate with the height of the shock layer. This depletion window effectively reduces the attenuation of a 4 GHz signal from 60\% to 4\% with a manageable power requirement of $66$~W per cm$^2$ of exposed cathode surface. {Feasibility analysis indicates that this system can be powered by a battery pack weighing less than 3 kg for a typical re-entry trajectory, with further mass reductions possible through intermittent transmission.} A sensitivity analysis reveals that the sheath topology is governed principally by ion kinetics; specifically, corrections to ion mobility at high reduced electric fields lead to enhanced space-charge shielding and a subsequent contraction of the sheath. Conversely, the sheath structure is largely insensitive to the electron mobility model. Finally, we argue that the present drift-diffusion model likely yields a conservative lower bound for mitigation performance. A kinetic approach accounting for ballistic ion transport and non-local ionization would likely predict thicker sheaths and lower attenuation for equivalent power deposition.
}

\nomenclature{
\begin{nomenclaturelist}{Roman symbols}

\item[$A$]{Arrhenius pre-exponential factor, cm$^3$/(mol-s-K$^n$) }
\item[$C_k$]{charge of $k$th  species, C}
\item[$C_{\rm i}$]{charge of ion species, C}
\item[$C_{\rm e}$]{charge of electrons, C}
\item[$c$]{speed of light, m/s}
\item[$(c_p)_k$]{specific heat at constant pressure of species $k$, J/kg-K}
\item[$d_{\rm p}$]{thickness of the plasma layer above the sheath, m}
\item[$d_{\rm s}$]{cathode sheath thickness, m}
\item[$E_{\rm i}$]{electric field along $i$th dimension, V/m}
\item[$\vec{E}$]{electric field vector, V/m}
\item[$E^\star$]{reduced electric field, $E/N$, V-m$^2$}
\item[$E_k^\star$]{reduced electric field of $k$th neutral species, V-m$^2$}
\item[$\mathcal{E}_{kl}$]{ activation energy of the $l$th electron impact process of the $k$th species, J/particle}
\item[$e_{\rm e}$]{electron specific internal energy, J/kg}
\item[$e_{\rm t}$]{total specific internal energy, J/kg}
\item[$e_{\rm v}$]{nitrogen average specific vibrational energy (excluding the zero-point energy), J/kg}
\item[$e_{\rm v}^0$]{nitrogen average specific vibrational energy in equilibrium (excluding the zero-point energy), J/kg}
\item[$f_{\rm p}$]{plasma frequency, Hz}
\item[$f_{\rm s}$]{signal frequency, Hz}
\item[$h_{k}^{0}$]{heat of formation of $k$th species, J/kg}
\item[$h_k$]{specific enthalpy of $k$th species excluding heat of formation, J/kg}
\item[$h_{\rm e}$]{electron specific enthalpy, J/kg}
\item[$I$]{transmitted signal intensity, W}
\item[$I_0$]{incident signal intensity, W}
\item[$\vec{J}$]{current density vector, A/m$^2$}
\item[$k_{\rm B}$]{Boltzmann constant, J/K}
\item[$k_{kl}$]{electron impact reaction rate of $l$th process on $k$th neutral species, $\rm m^3/s$}
\item[$\rm M_{\infty}$]{freestream Mach number}
\item[$m_k$]{particle mass of $k$th species, kg}
\item[$m_{\rm e}$]{electron mass, kg}
\item[$N$]{number density of gas mixture, $\rm m^{-3}$}
\item[$N_{\rm e}$]{number density of electrons, $\rm m^{-3}$}
\item[$N_{\rm i}$]{number density of ions, $\rm m^{-3}$}
\item[$N_k$]{number density of $k$th species, $\rm m^{-3}$}
\item[$N_{\rm p}$]{average number density of electrons in the plasma above the sheath, $\rm m^{-3}$}
\item[$n$]{Arrhenius temperature exponent}
\item[$n_{\rm s}$]{number of species}
\item[$P_{\rm e}$]{electron pressure, Pa}
\item[$P_k$]{partial pressure of $k$th species, Pa}
\item[$P$]{pressure of the gas mixture, $\sum P_k$, Pa}
\item[$P_{\rm dyn}$]{dynamic pressure, Pa}
\item[$P_{\infty}$]{freestream pressure, Pa}
\item[$\cal P$]{power deposited to flow per cathode surface area, W/cm$^2$}
\item[$Q_{\rm e-i}$]{electron cooling rate per unit volume to all inelastic collisions, W/m$^3$ }
\item[$Q_{\rm e-t} $]{electron cooling-heating rate per unit volume due to elastic collisions, W/m$^3$ }
\item[$Q_{\rm e-v} $]{electron cooling rate per unit volume due to electron-vibrational collisions, W/m$^3$ }
\item[$Q_{\rm v-e} $]{electron heating rate per unit volume due to electron-vibrational collisions, W/m$^3$ }
\item[$\overline{{q}_{\rm e}}$]{electron thermal speed, m/s}
\item[$q_\infty$]{freestream flow speed, m/s}
\item[$R$]{leading edge nose cap radius, mm}
\item[$\cal R$]{universal gas constant, cal/K-mol}
\item[$S$]{cathode surface area, cm}
\item[$s_k$]{sign of the charge of species $k$}
\item[$T_{\infty}$]{freestream temperature, K}
\item[$T$]{bulk gas temperature, K}
\item[$T_{\rm e}$]{electron temperature, K}
\item[$T_{\rm ref}$]{reference temperature in electron swarm experiments, K}
\item[$T_{\rm v}$]{vibrational temperature, K}
\item[$T_{\rm wall}$]{wall temperature, K}
\item[$\cal T$]{pulse period, s}
\item[$t$]{time, s}
\item[$u_{d}$]{drift velocity, m/s}
\item[$V_{i}$]{bulk flow velocity component, m/s}
\item[$V_{i}^k$]{component of $k$th species velocity, m/s}
\item[$\vec{V}_{\rm e}$]{electron velocity vector, m/s}
\item[$\Vol$]{volume of computational domain, m$^3$}
\item[$W_{\rm e}$]{chemical source term of electrons, kg/(m$^3$-s)}
\item[$W_k$]{chemical source term of $k$th species, kg/(m$^3$-s) }
\item[$W_{\rm N_2}$]{chemical source term of nitrogen, kg/(m$^3$-s) }
\item[$w_k$]{mass fraction of $k$th species}
\item[$w_{\rm N_2}$]{nitrogen mass fraction}
\item[$x_{i}$]{Cartesian coordinate, m}
  \end{nomenclaturelist}

\begin{nomenclaturelist}{Greek symbols}
\item[$\beta_{k}^{\rm i}$]{parameter equal to 1 when $k$th species is ionic, 0 otherwise}
\item[$\beta_{k}^{\rm n}$]{parameter equal to 1 when $k$th species is neutral, 0 otherwise}
\item[$\beta_{k}^{+}$]{parameter equal to 1 when $k$th species is a positive ion, 0 otherwise}
\item[$\gamma$]{surface catalytic recombination coefficient}
\item[$\gamma_{\rm e}$]{electron secondary emission coefficient}
\item[$\epsilon_0$]{permittivity of free space, C/(V-m)}
\item[$\epsilon_r$]{relative permittivity}
\item[$\zeta_{\rm v}$]{fraction of electron energy consumed in the excitation of the nitrogen vibrational energy}
\item[$\eta$]{viscosity, Pa-s}
\item[$\theta_{\rm v}$]{characteristic nitrogen vibration temperature, K}
\item[$\kappa_{\rm e}$]{electron thermal conductivity, W/(m-K)}
\item[$\kappa_{\rm i}$]{ion thermal conductivity, W/(m-K)}
\item[$\kappa_{\rm n}$]{neutrals thermal conductivity, W/(m-K)}
\item[$\kappa_{\rm v}$]{nitrogen vibrational thermal conductivity, W/(m-K)}
\item[$\ln \Lambda$]{Coulomb logarithm}
\item[$\mu_{\rm e}$]{electron mobility, m$^2$/(V-s)}
\item[$(\mu_{\rm e}^\star)_k$]{reduced electron mobility in $k$th neutral species, $(\mu_{\rm e}N)_k$, $\rm m^{-1}V^{-1}s^{-1}$}
\item[$\mu_{\rm i}$]{ion mobility, m$^2$/(V-s)}
\item[$\mu_k$]{mobility of $k$th charged species, m$^2$/(V-s)}
\item[$\nu_{\rm en}$]{electron-neutral collision frequency, 1/s}
\item[$\nu_k$]{mass diffusion coefficient of $k$th species, kg/(m-s)}
\item[$\nu_{\rm N_2}$]{mass diffusion coefficient of nitrogen, kg/(m-s)}
\item[$\xi$]{mobility weighting factor}
\item[$\rho$]{mass density of the mixture, $\sum_k \rho_k$, kg/m$^3$}
\item[$\rho_c$]{net charge density, C/m$^3$}
\item[$\rho_{\rm e}$]{electron mass density, kg/m$^3$}
\item[$\rho_{k}$]{partial mass density of $k$th species, kg/m$^3$}
\item[$\rho_{\rm N_2}$]{nitrogen mass density, kg/m$^3$}
\item[$\tau_{ji}$]{viscous stress tensor, Pa}
\item[$\tau_{\rm vt}$]{vibrational-translational relaxation time, s}
\item[$\phi$]{electric field potential, V}
\item[$\phi_{\rm s}$]{voltage drop in cathode sheath, V}
\item[$\chi$]{wall-normal coordinate positive towards fluid, m}
\item[$\omega_{\rm p}$]{plasma circular frequency, 1/s}
\item[$\omega_{\rm s}$]{signal circular frequency, 1/s}
\end{nomenclaturelist}

}

\begin{document}
\maketitle
\makenomenclature

\section{Introduction}

\dropword Hypersonic reentry into Earth's atmosphere normally occurs at speeds above 6 km/s. The shock-heated ambient air becomes ionized and envelops the vehicle in a thick plasma layer that extends in its wake. When the density of this plasma layer is high ($10^{16}-10^{19}~{\rm electrons/m^3}$) it can reflect, absorb and scatter incoming electromagnetic (EM) waves, disrupting communication with the vehicle, a mission-critical phenomenon known as ``communications blackout'' that can last several minutes. The plasma will be opaque to incoming frequencies that are below the plasma frequency, an increasing function of the plasma number density.

The severity of attenuation depends strongly on the frequency band: L--band and S--band links (1–2 GHz and 2–4 GHz) are most susceptible, often experiencing complete signal loss when the local plasma frequency exceeds the carrier frequency, while C--band and X--band systems (4–8 GHz and 8–12 GHz) can retain partial link capability but still suffer measurable degradation. These trends were observed in the RAM-C-I and RAM-C-II flight experiments [\cite{nasa:1970:grantham}] between altitudes of 60 to 80 kilometers, where telemetry showed near-total attenuation of L-band signals, significant S-band degradation, and improved but still degraded performance at C--band and X--band. The RAM-C-II electron density data remains a cornerstone for validating modern reentry plasma models.

Several methods have been proposed to alleviate this blackout problem. First, aerodynamic shaping, described in \cite{nasa::1994:gillman}, aims to modify the vehicle's airflow to disrupt or minimize the formation of a dense plasma, using slender and sharp-edge geometries to reshape the shockwave but with potential disadvantages for aerodynamic performance and heating loads. Alternatively, magnetic-window approaches by \cite{jap:2007:korotkevich} or \cite{aiaaconf:2025:anderson} aim to create new propagation modes in the plasma by applying magnetic fields and make it appear ``transparent'' to incoming signals. However, variable magnetic fields may be required as the reentry plasma density varies throughout the reentry trajectory, leading to weight constraints and limited by advancements in magnetic materials. Approaching mitigation through chemical kinetics, a feed system based on water or electrophilic injectants may also be used to reduce plasma density by increasing recombination, despite the design complexity [\cite{nasa::1968:schroeder, aiaaconf:2023:sawicki}]. Finally, simplified schemes for the manipulation of the plasma layer by MHD methods have been assessed by \cite{jsr:2010:kim} and \cite{rast:2011:kim} employing crossed electric and magnetic fields to accelerate electrons away from suitable antenna locations.

The design complexity or payload budget penalties associated with the above approaches may be avoided by only using electric fields produced by high negative voltages as suggested by \cite{jsr:2008:keidar}. Applying such a field leads to the generation of a large non-neutral plasma sheath near the cathode, characterized by an electron number density several orders of magnitude lower than the ion density. Because this sheath can be quite thick and electromagnetic wave interference is primarily determined by electron density rather than ion density, its formation significantly reduces signal attenuation. The effect of pulsed discharges has been studied experimentally by \cite{psst:2024:luo} showing a reduction of 86\% in electron density in the actuator region. Particle--In--Cell (PIC) simulations by \cite{ieee:2017:steffens} using pulsed discharges showed a electron depletion region spanning a few tens of the Debye length using fast-switching ($\sim$10 ns) pulses in an argon plasma. Further work by \cite{jpd:2017:krishnamoorthy} verified the feasibility of this approach with PIC simulations using constant plasma parameters.

Prior work on the use of an electric field to reduce electron density has focused on simulations with spatially-fixed constant plasma parameters and without accounting for fully-coupled interactions. In this context, fully-coupled phenomena arises when the  power deposited to the gas by the electric field is sufficiently high to alter the gas properties (temperature, density) significantly.  

Coupling the Navier-Stokes equations for bulk flow with a drift-diffusion model for charged species and a Gauss-law-based potential equation for electric field is computationally challenging, primarily due to the disparate time scales of neutrals and charged species. This discrepancy typically results in a stiff system, historically attributed to stability constraints imposed by the smallest scales that necessitate prohibitively small time steps. Consequently, capturing the largest time scales required excessive iterations, often forcing the use of coarse meshes that compromised accuracy. However, \cite{jcp:2014:parent,aiaa:2016:parent} demonstrated that for block-implicit schemes, this stiffness arises not from time scale discrepancies, but from error amplification within the Gauss-based potential equation. To mitigate this, the electric field can instead be derived from Ohm's law, with source terms added to the ion transport equations to enforce Gauss's law. This reformulation allows non-neutral sheaths to be integrated using aerodynamic-scale time steps, significantly improving computational efficiency and reducing numerical error in fully coupled simulations.

In this work, we employ the advanced numerical framework developed by \cite{jcp:2014:parent,aiaa:2016:parent} and summarized in \cite{book:2022:parent} to simulate, for the first time, the interaction of pulsed discharges with the plasma flow around a re-entry vehicle. Starting from a steady-state solution at Mach 24, we apply triangle-waveform pulses to the vehicle's wedge section, inducing a region of electron depletion above the cathode {that effectively reduces signal attenuation.} {Beyond demonstrating mitigation performance, we isolate the critical roles of ion mobility and secondary electron emission in governing sheath topology. Finally, we establish why the present drift-diffusion approximation likely yields a conservative lower bound for performance compared to a fully kinetic description.}

The paper is organized as follows: Sections II and III detail the physical model and numerical methods, respectively. Section IV validates the model against planetary entry flight tests and experimental data{.} {Sections V and VI cover the problem setup and grid convergence, while Section VII defines the performance metrics. Finally, Section VIII analyzes the simulation results, detailing the physical mechanisms driving the depletion layer.}

\section{Physical Model}

The motion of charged particles with respect to the bulk is obtained through the drift-diffusion model, {following \cite{book:2022:parent}}. Thus, for the electrons and ions, the velocity difference with respect to the bulk involves both a drift and a diffusion component. For
the neutrals, the velocity difference with respect to the bulk is set proportional to the product between the mass fraction gradient and the mass diffusion coefficient:
\begin{equation}
  V^{k}_i = \left\{
  \begin{array}{ll}\mfd
  V_i+s_k \mu_k  {E}_i
             -  \frac{\mu_k}{|C_k| N_k} \frac{\partial P_k}{\partial x_i} & \textrm{for charged species} \alb\mfd
  V_i - \frac{\nu_k}{\rho_k} \frac{\partial w_k}{\partial x_i} & \textrm{for neutral species}
  \end{array}
  \right.
\label{eqn:Vk}
\end{equation}
where $V^{k}_i$ is the species velocity, $V_i$ is the bulk velocity of the plasma, $E$ is the electric field, $\mu_k$ is the mobility, $N_k$ is the species number density and $P_k$ is the partial pressure. As well, $C_k$ is the elementary charge of single-charge species and $s_k$ is the sign of the charge of the species (-1 for electrons or negative ions, +1 for positive ions). The velocity $V^{k}_i$ of the neutrals is written in terms of the diffusion term with respect to the bulk, where $\nu_k$ is the mass diffusion coefficient of neutral species $k$, $\rho_k$ is the mass density and $w_k$ is the mass fraction of species $k$. {As done in \cite{nasa:1989:gnoffo} for each $k$th  species, either charged or neutral, we can write the mass conservation equation as follows:}
\begin{equation}
\frac{\partial}{\partial t} \rho_k + \sum_i \frac{\partial }{\partial x_i}\rho_{k} V_i^{k} = W_{k}  
\label{eqn:masstransport}
\end{equation}
The momentum equation for the bulk flow corresponds to the Navier-Stokes equations {as outlined in \cite{aiaa:2016:parent},} taking into account the force due to an electric field:
\begin{equation}
  \rho \frac{\partial V_i }{\partial t}+ \sum_{j=1}^3 \rho V_j \frac{\partial V_i}{\partial x_j}
=
-\frac{\partial P}{\partial x_i} 
+ \sum_{j=1}^3 \frac{\partial \tau_{ji}}{\partial x_j}
+ \rho_{\rm c}{E}_i
\label{eqn:momentumtransport}
\end{equation}
Here, we consider only the vibrational energy transport of N$_2$ and assume that the vibrational temperature of O$_2$ and NO matches the one of the bulk gas temperature. The nitrogen vibrational energy transport equation, {taken from \cite{nasa:1989:gnoffo}} is:
\begin{equation}
 \begin{array}{l}
  \mfd\frac{\partial}{\partial t} \rho_{\rm N_2} e_{\rm v}
     + \sum_{j=1}^{3} \frac{\partial }{\partial x_j}
       \rho_{\rm N_2} V_j e_{\rm v}
     - \sum_{j=1}^{3} \frac{\partial }{\partial x_j} \left(
            \kappa_{\rm v}  \frac{\partial T_{\rm v}}{\partial x_j}\right)\alb\mfd
     - \sum_{j=1}^{3} \frac{\partial }{\partial x_j} \left(
            e_{\rm v} \nu_{\rm N_2}  \frac{\partial w_{\rm N_2}}{\partial x_j}\right)
 = 
Q_{\rm e-v}-Q_{\rm v-e}   + \frac{\rho_{\rm N_2}}{\tau_{\rm vt}}\left( e_{\rm v}^0 -e_{\rm v} \right) + W_{\rm N_2} e_{\rm v}
\end{array}
\label{eqn:vibrationalenergytransport}
\end{equation}
where $e_{\rm v}$ , $e_{\rm v}^0$ and $\kappa_{\rm v}$ are the nitrogen vibrational energy, vibrational energy in equilibrium and the nitrogen vibrational thermal conductivity, respectively. As well, $Q_{\rm e-v}-Q_{\rm v-e}$ is the net energy transfer from the electrons to the nitrogen vibrational energy modes outlined in  \cite{pf:2025:rodriguez2} and $\tau_{\rm vt}$ is the vibrational-translational relaxation time from \cite{nasa:1989:gnoffo}. 

The electron energy transport equation can be derived from the first law of thermodynamics applied to electrons following \cite{book:1991:raizer} and neglecting electron inertia terms:
\begin{align}
 \frac{\partial }{\partial t}\rho_{\rm e} e_{\rm e} &+ \sum_{j=1}^3  \frac{\partial }{\partial x_j} \left( \rho_{\rm e} V^{\rm e}_j h_{\rm e}  - \kappa_{\rm e} \frac{\partial T_{\rm e}}{\partial x_j}  
\right)\nonumber \alb
&= 
 W_{\rm e} e_{\rm e}
+   C_{\rm e} N_{\rm e} \vec{E} \cdot \vec{V}_{\rm e}
 -Q_{\rm e-t}-Q_{\rm e-i}+Q_{\rm v-e}
\label{eqn:electronenergytransport}
\end{align}
In the electron energy transport equation, a critical source term that requires adequate modeling is the rate of electron energy loss due to inelastic collisions. This leads to electron cooling and can be expressed first in standard form [see for instance  \cite{jap:2019:peters} or \cite{jap:2023:pokharel}]:
\begin{equation}
Q_{\rm e-i}  
=   \sum_k   \beta_k^{\rm n} N_{\rm e} N_k  \sum_l k_{kl} \mathcal{E}_{kl}
\label{eqn:Qinelastic_standard}
\end{equation}
where $N_k$ is the $k$th neutral species number density, $N_{\rm e}$ is the electron number density, $l$ denotes an electron impact process, $k_{kl}$ is the rate coefficient of the $l$th electron impact process acting on the $k$th neutral species, $\mathcal{E}_{kl}$ is the activation energy of the $l$th electron impact process of the $k$th species and $\beta_k^{\rm n}$ is 1 when the $k$th species is a neutral and to 0 otherwise. Following \cite{pf:2024:parent}, the electron energy loss rate in inelastic collisions can be shown to correspond to:

\begin{equation}
  \sum_l k_{kl} \mathcal{E}_{kl}  
=  |C_{\rm e}|  \left(\mu_{\rm e}^\star\right)_k \left( (E^\star_k)^2 - \frac{3  k_{\rm B}    (T_{\rm e}-T_{\rm ref})}{ m_{k} (\mu_{\rm e}^\star)^2_k}\right) 
\label{eqn:suminelasticenergies}
\end{equation}
{The latter is written} in terms of the $k$th species reduced electric fields $E^\star_k$ taken from \cite{pf:2025:rodriguez2} and reduced electron mobilities in the $k$th species $\left(\mu_{\rm e}^\star\right)_k$ outlined in \cite{pf:2024:parent}. Substituting the latter in the former, the total inelastic electron cooling rate becomes:
\begin{equation}
Q_{\rm e-i}  
=  \sum_k \beta_k^{\rm n} |C_{\rm e}| N_{\rm e} N_k (\mu_{\rm e}^\star)_k \left( (E^\star_k)^2 -  \frac{3  k_{\rm B}    (T_{\rm e}-T_{\rm ref})}{ m_{k} (\mu_{\rm e}^\star)^2_k} \right) 
\label{eqn:Qinelastic_cooling}
\end{equation}
To find the energy transfer from the electrons to the vibrational modes of nitrogen in the vibrational energy transport equation, we note that this corresponds to the product between the fraction of electron energy loss $\zeta_{\rm v}$ given in \cite{pf:2025:rodriguez2}, consumed in the excitation of the nitrogen vibrational energy, and the total electron cooling due to all inelastic $\rm N_2$ processes:
\begin{equation}
Q_{\rm e-v} =   \zeta_{\rm v} |C_{\rm e}| N_{\rm e} N_{\rm N_2} (\mu_{\rm e}^\star)_{\rm N_2} \left( (E^\star_{\rm N_2})^2 -  \frac{3  k_{\rm B}    (T_{\rm e}-T_{\rm ref})}{ m_{\rm N_2} (\mu_{\rm e}^\star)^2_{\rm N_2}} \right) 
\label{eqn:Q_ev}
\end{equation}
The $\rm N_2$ vibrational-electron heating rate $Q_{\rm v-e}$ can be obtained from the electron-vibrational cooling rate following \cite{pf:2025:rodriguez2}:
\begin{align}
\frac{Q_{\rm v-e}}{Q_{\rm e-v}}
&= 
 \exp \left(\dfrac{\theta_{\rm v}}{T_{\rm e}}-\dfrac{\theta_{\rm v}}{T_{\rm v}}\right) 
\label{eqn:scale_proposed}
\end{align}
where the characteristic nitrogen vibration temperature $\theta_{\rm v}$ corresponds to 3353~K, as suggested in \cite{book:1962:barrow}. The electron cooling-heating due to elastic collisions, which we also take from \cite{pf:2024:parent} corresponds to:
\begin{align}
Q_{\rm e-t}
&= 
   \underbrace{\sum_k  \frac{3 \beta_k^{\rm n} k_{\rm B} |C_{\rm e}| N_{\rm e} N_k  (T_{\rm e}-T)}{ m_k (\mu_{\rm e}N)_k}}_{\rm elastic~cooling-heating~to~neutrals}\nonumber\\
 &+ \underbrace{\sum_k \beta_k^{\rm i}  N_{\rm e} N_k  (T_{\rm e}-T )  \frac{6 k_{\rm B} C_{\rm i}^2 C_{\rm e}^2 \ln \Lambda}{ \pi^3 \epsilon_0^2 m_{\rm e} m_k \overline{q_{\rm e}}^3}}_{\rm elastic~cooling-heating~to~ions} 
\label{eqn:Qelastic}
\end{align}
with $\ln \Lambda$ the recommended Coulomb logarithm which can be found in \cite[page 34]{nrl:2002:huba} and $(\mu_{\rm e} N)_k\equiv (\mu_{\rm e}^\star)_k$ corresponds to the reduced electron mobility of the $k$th neutral species. The translational temperature $T$ is determined from the total energy equation:
\begin{equation}
\begin{array}{l}\mfd
 \frac{\partial }{\partial t}\rho e_{\rm t}
+ \sum_{j=1}^3  \frac{\partial }{\partial x_j} V_j \left(\rho  e_{\rm t} +  P \right)
 \alb\mfd
- \sum_{j=1}^3  \frac{\partial }{\partial x_j} \left(
   \nu_{\rm N_2} e_{\rm v}\frac{\partial w_{\rm N_2}}{\partial x_j} 
  + \sum_{k=1}^{\rm n_s}  \rho_k (V^k_j-V_j) {(h_k+h_k^\circ)}
\right)
 \alb\mfd
-\sum_{i=1}^{3}\frac{\partial }{\partial x_i}\left((\kappa_{\rm n}+\kappa_{\rm i}) \frac{\partial T}{\partial x_i} 
+ \kappa_{\rm v} \frac{\partial T_{\rm v}}{\partial x_i} +\kappa_{\rm e} \frac{\partial T_{\rm e}}{\partial x_i}\right)
 \alb\mfd
=
 \sum_{i=1}^3 \sum_{j=1}^3  \frac{\partial }{\partial x_j} \tau_{ji} V_i
+ \vec{E}\cdot\vec{J}
\end{array}
\label{eqn:totalenergytransport}
\end{equation}
where $h_k^\circ$ is the heat of formation and the sum $(h_k+h_k^\circ)$ represents the enthalpy of the $k$th species including calorically-imperfect effects as well as the heat of formation, obtained from the high temperature enthalpy polynomials by \cite{nasa:2002:mcbride}. As well, $\vec{J} \equiv \sum_{k}  C_k N_k  \vec{V}_k$ is the current density vector and $e_{\rm t}$ is the total specific energy. 

{The total energy equation outlined above is obtained by summing the energy equations for the ions, electrons and neutrals. For each species, the energy equation is obtained from the first law of thermodynamics and the momentum equation (either drift--diffusion or Navier-–Stokes) applied to each species. Thus, in our definition of the total energy $e_{\rm t}$ we do not include the kinetic energy of the electrons or ions because the charged species momentum equations based on the drift-diffusion model do not include the inertia terms.} 

Closure of the latter system of transport equations requires {the enforcement of} Gauss's law as follows:
\begin{equation} 
\sum_{i=1}^3 \frac{\partial }{\partial x_i}\left(\epsilon_r \frac{\partial \phi}{\partial x_j}\right)=-\frac{\rho_{\rm c}}{\epsilon_0}   \label{eqn:gauss_potential} 
\end{equation}
with $\rho_c$, $\phi$, $\epsilon_0$, and $\epsilon_{\rm r}$ the net charge density, the electric field potential, the permittivity of free space, and the relative permittivity, respectively. {Details on the form of Eq. (\ref{eqn:gauss_potential}) can be consulted in \cite{jcp:2014:parent}.} From the potential we can find the electric field by taking the negative of its gradient (i.e.\ $\vec{E}=-\vec{\nabla} \phi$). Since there is no externally applied magnetic field and the magnetic Reynolds number is low, it is sufficient to obtain an electrostatic solution of the electric potential from Eq.~(\ref{eqn:gauss_potential}) rather than solving the complete set of Maxwell's equations of electromagnetism.

Unless indicated otherwise, the viscosity of the gas mixture, species thermal conductivities, mobilities and other transport coefficients are taken from the \cite{nasa:1990:gupta} high-temperature transport model. The charged species mobilities are determined from Einstein–Smoluchowski relation from the binary diffusion coefficients. We use the correction outlined in \cite{jtht:2023:parent} in determining the thermal conductivities of the charged species:
\begin{equation}
  \kappa_k = \frac{\rho_k k_{\rm B} (c_p)_k T_k \mu_k}{|C_k|}    
\end{equation}
where $(c_p)_k$ is the specific heat at constant pressure of the $k$th species and $T_k$ is the species temperature (either $T$ for the ions or $T_{\rm e}$ for the electrons). Again following \cite{jtht:2023:parent}, a second modification to the Gupta-Yos model we implement here is to exclude the electron-electron collisions when determining the electron mobilities.

We note that the \cite{nasa:1990:gupta} transport model uses collision integrals function of temperature that are valid up to 30,000 K. While the temperature of ions (the bulk gas temperature) is well below this value, the electron temperature can reach hundreds of thousands of Kelvin in discharge regions. For this reason, when presenting the results in this paper, we will rather use expressions for the reduced mobility of electrons (function of electron temperature) in each neutral species obtained with BOLSIG+ or swarm experiments as outlined in \cite{pf:2024:parent}.

At the solid boundaries, we assume no surface catalysis for neutral species. For charged species it is assumed that electrons and ions recombine fully. We here model secondary electron emission (SEE) through the SEE coefficient $\gamma_{\rm e}$. This results in the following boundary condition for the electron number density when the electric field points towards the surface:
\begin{equation}
N_{\rm e}=\frac{\gamma_{\rm e}}{\mu_{\rm e}} \sum_{k=1}^{n_{\rm s}} N_k \mu_k \beta_k^+
{~~\rm for~}
E_\chi<0
\end{equation}
and in the following when the electric field points away from the surface:
\begin{equation}
\frac{\partial }{\partial \chi} N_{\rm e} V^{\rm e}_\chi= 0
{~~\rm for~}
E_\chi>0
\end{equation}
In the latter, $\beta_k^+$ is 1 when the $k$th species is a positive ion and 0 otherwise. As well, $\chi$ is a coordinate normal to the boundary and positive pointing to the plasma. While the above boundary condition is written in terms of the positive ions, we note that $\gamma_{\rm e}$ should include the effect of ion-electron, electron-electron, and neutral-electron collisions at the surface. Therefore, the effective secondary electron yield per ion can be higher than if it accounted only for ion-electron collisions at the electrode surface. The effect of the choice of $\gamma_{\rm e}$ for the electrodes will be assessed in the discussion of the results. As for positive ions, their density is fixed at zero when the electric field points surface-ward, while a zero-gradient condition applies in the reverse case. 

Chemical kinetics are obtained for 11 air species $\rm N_2$, $\rm O_2$, $\rm NO$, $\rm N$, $\rm O$, $\rm N_2^+$, $\rm O_2^+$, $\rm NO^+$, $\rm N^+$, $\rm O^+$ and $\rm e^-$ with the reaction list outlined in Table~\ref{tab:chemical_model}. For the dissociation reactions 1, 2 and 4--7 we use the forward control temperature $\sqrt{TT_{\rm v}}$ recommended by \cite{jtht:1993:park}. This control temperature is based on the preferential dissociation concept, where dissociation of molecular species occurs more easily when the molecules are vibrationally excited. As such, the control temperature is weighed heavily by the vibrational temperature.

\begin{figure}[ht]
    \centering
     \includegraphics[width=0.40\textwidth]{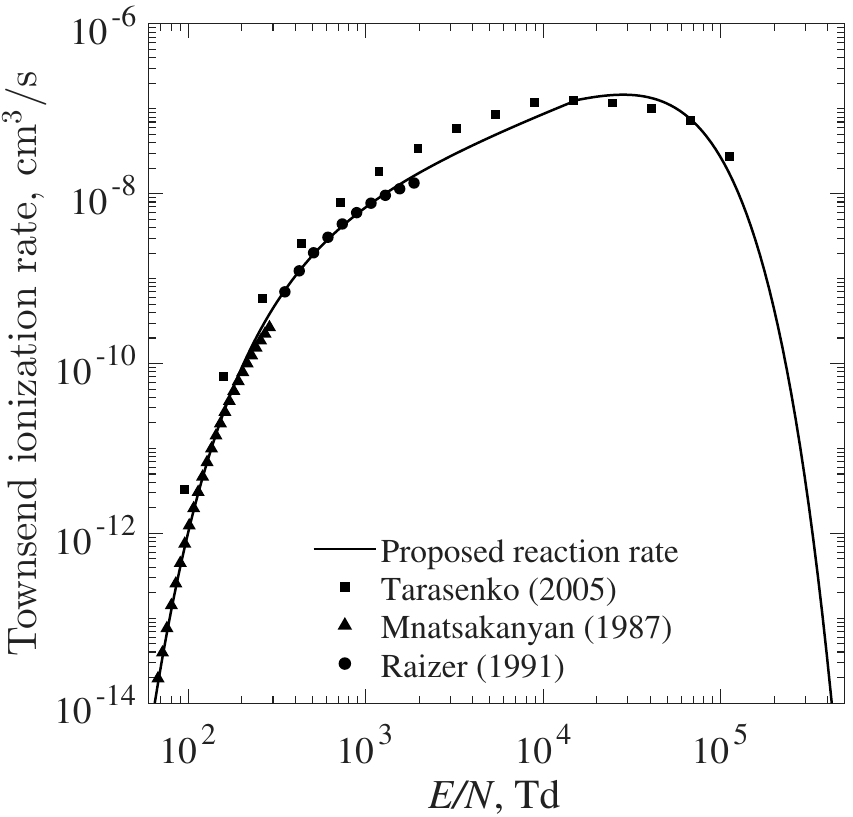}
     \figurecaption{Comparison of proposed $\rm N_2$ Townsend ionization rate with available experimental data.}
     \label{fig:townsendrates}
\end{figure}

In high $E/N$ field discharge regions such as the cathode sheaths, Townsend ionization rates (reactions 24 and 25) are expressed as functions of the reduced electric field $E^\star \equiv |E/N|$ because such are well tabulated from experiments while proper modeling of electron temperature in this range is not well known. The proposed Townsend ionization rate for nitrogen, valid over a wide range of $E/N$, is shown in Fig.~\ref{fig:townsendrates} and compared with data from \cite{ps:2005:tarasenko}, 
\cite{kp:1987:mnatsakanyan}, and \cite{book:1991:raizer}. The ionization rate for $\rm O_2$ is obtained in a similar fashion from the rate for air and for $\rm N_2$  given in \cite{book:1991:raizer} assuming a $\rm N_2$:$\rm O_2$ ratio of 4:1.

\begin{table*}[!t]
  \center
  \begin{threeparttable}
\tablecaption{Reaction mechanism and rate coefficients for the 11-species high-temperature air model.\tnote{a,b,c}}
    \label{tab:chemical_model}
    \fontsizetable
    \begin{tabular*}{\textwidth}{@{}l@{\extracolsep{\fill}}c@{~~~}cccccccc@{}}
    \toprule
~&~&\multicolumn{4}{c}{Forward rate} & \multicolumn{4}{c}{Backward rate} \\
 \cmidrule(lr){3-6}\cmidrule(lr){7-10}
No.~~~&{Reaction} & $T$ & $A$ & $n$ & $E$ & $T$  & $A$ & $n$ & $E$ \\ 
\midrule
1 & $\rm N_2 + M_1 \rightleftarrows N + N + M_1$ & $\sqrt{TT_{\rm v}}$   & 3.0 $\cdot$ 10$^{22}$  & $-1.6$ & $113200 \, {\cal R}$  & $T$ &\multicolumn{3}{c}{Equilibrium constant} \\
2 & $\rm N_2 + M_2 \rightleftarrows N + N + M_2$ &$\sqrt{TT_{\rm v}}$  & 7.0 $\cdot$ 10$^{21}$  & $-1.6$ & $113200 \, {\cal R}$ & $T$ & \multicolumn{3}{c}{Equilibrium constant}  \\

3 & $\rm N_2 + e^- \rightleftarrows N + N + e^-$ & $T_{\rm e}$ & 3.0 $\cdot$ 10$^{24}$  & $-1.6$ & $113200 \, {\cal R}$ & $T_{\rm e}$ & \multicolumn{3}{c}{Equilibrium constant} \\

4 & $\rm O_2 + M_1 \rightleftarrows O + O + M_1$ & $\sqrt{TT_{\rm v}}$  & 1.0 $\cdot$ 10$^{22}$  & $-1.5$ & $59500 \, {\cal R}$ & $T$ & \multicolumn{3}{c}{Equilibrium constant} \\

5 & $\rm O_2 + M_2 \rightleftarrows O + O + M_2$ & $\sqrt{TT_{\rm v}}$  & 2.0 $\cdot$ 10$^{21}$  & $-1.5$ & $59500 \, {\cal R}$ & $T$ & \multicolumn{3}{c}{Equilibrium constant} \\

6 & $\rm NO + M_3 \rightleftarrows N + O + M_3$ & $\sqrt{TT_{\rm v}}$  & 1.1 $\cdot$ 10$^{17}$  & 0.0 & $75500 \, {\cal R}$ & $T$ & \multicolumn{3}{c}{Equilibrium constant} \\

7 & $\rm NO + M_4 \rightleftarrows N + O + M_4$ &$\sqrt{TT_{\rm v}}$  & 5.0 $\cdot$ 10$^{15}$  & 0.0 & $75500 \, {\cal R}$ & $T$ & \multicolumn{3}{c}{Equilibrium constant}  \\

8 & $\rm NO + O \rightleftarrows N + O_2 $ & $T$  & 8.4 $\cdot$ 10$^{12}$  & 0.0 & $19400 \, {\cal R}$ & $T$ & \multicolumn{3}{c}{Equilibrium constant}  \\

9 & $\rm N_2 + O \rightleftarrows NO + N $ & $T$  & 5.7 $\cdot$ 10$^{12}$  & $0.42$ & $42938 \, {\cal R}$ & $T$ & \multicolumn{3}{c}{Equilibrium constant} \\

10 & $\rm N + O \rightleftarrows NO^+ + e^- $ & $T$  & 5.3 $\cdot$ 10$^{12}$  & 0.0 & $32000 \, {\cal R}$ & $T_{\rm e}$ & $3.00 \cdot 10^{-7} \cdot {\cal A} \cdot300^{0.56}$ & $-0.56$ & $0.0$ \\

11 & $\rm O + O \rightleftarrows O_2^+ + e^- $ & $T$  & 1.1 $\cdot$ 10$^{13}$  & 0 & $81200 \, {\cal R}$ & $T_{\rm e}$ & $2.40 \cdot 10^{-7} \cdot {\cal A}\cdot300^{0.70} $ & $-0.70$ & $0.0$\\

12 & $\rm N + N \rightleftarrows N_2^+ + e^- $ & $T$ & 2.0 $\cdot$ 10$^{13}$  & 0 & $67700 \, {\cal R}$ & $T_{\rm e}$ &  \multicolumn{3}{c}{Rate data from Fig.~3 in \cite{aip:1998:peterson}} \\

13 & $\rm NO^+ + O \rightleftarrows N^+ + O_2 $ & $T$   & 1.0 $\cdot$ 10$^{12}$  & 0.5 & $77200 \, {\cal R}$ & $T$& \multicolumn{3}{c}{Equilibrium constant}\\

14 & $\rm N^+ + N_2 \rightleftarrows N_2^+ + N $ & $T$   & 1.0 $\cdot$ 10$^{12}$  & 0.5 & $12200 \, {\cal R}$ & $T$ & \multicolumn{3}{c}{Equilibrium constant}\\

15 & $\rm O_2^+ + N \rightleftarrows N^+ + O_2 $ & $T$   & 8.7 $\cdot$ 10$^{13}$  & 0.14 & $28600 \, {\cal R}$ & $T$ & \multicolumn{3}{c}{Equilibrium constant}\\

16 & $\rm O^+ + NO \rightleftarrows N^+ + O_2 $ & $T$  & 1.4 $\cdot$ 10$^{5}$  & 1.90 & $26600 \, {\cal R}$ & $T$  & \multicolumn{3}{c}{Equilibrium constant}\\

17 & $\rm O_2^+ + N_2 \rightleftarrows N_2^+ + O_2 $ & $T$   & 9.9 $\cdot$ 10$^{12}$  & 0.00 & $40700 \, {\cal R}$ & $T$ & \multicolumn{3}{c}{Equilibrium constant}\\

18 & $\rm O_2^+ + O \rightleftarrows O^+ + O_2 $ & $T$   & 4.0 $\cdot$ 10$^{12}$  & $-0.09$ & $18000 \, {\cal R}$ & $T$ & \multicolumn{3}{c}{Equilibrium constant}\\

19 & $\rm NO^+ + N \rightleftarrows O^+ + N_2 $ & $T$   & 3.4 $\cdot$ 10$^{13}$  & $-1.08$ & $12800 \, {\cal R}$ & $T$ & \multicolumn{3}{c}{Equilibrium constant}\\

20 & $\rm NO^+ + O_2 \rightleftarrows O_2^+ + NO $ & $T$   & 2.4 $\cdot$ 10$^{13}$  & 0.41 & $32600 \, {\cal R}$ & $T$ & \multicolumn{3}{c}{Equilibrium constant}\\

21 & $\rm NO^+ + O \rightleftarrows O_2^+ + N $ & $T$   & 7.2 $\cdot$ 10$^{12}$  & 0.29 & $48600 \, {\cal R}$ & $T$ & \multicolumn{3}{c}{Equilibrium constant}\\

22 & $\rm O^+ + N_2 \rightleftarrows N_2^+ + O $ & $T$  & 9.1 $\cdot$ 10$^{11}$  & 0.36 & $22800 \, {\cal R}$ & $T$  & \multicolumn{3}{c}{Equilibrium constant}\\

23 & $\rm NO^+ + N \rightleftarrows N_2^+ + O $  & $T$  & 7.2 $\cdot$ 10$^{13}$  & 0.00 & $35500 \, {\cal R}$ & $T$ & \multicolumn{3}{c}{Equilibrium constant}\\

24 & $\rm O_2 + e^- \rightleftarrows O_2^+ + e^- + e^-$ &\multicolumn{4}{c}{$\min\left(k_{\rm i},~{\rm exp}(-0.0102785\cdot {\rm ln}^2 E^\star - 2.42260\cdot 10^{-75} \cdot {\rm ln}^{46}E^\star)\right)$} 
                                          &$T_{\rm e}$ & $2.2 \cdot 10^{40}$  & $-4.5$  & 0.0\\
25~ & $\rm N_2 + e^- \rightleftarrows N_2^+ + e^- + e^-$ &\multicolumn{4}{c}{$\min\left(k_{\rm i},~{\rm exp}(-0.0105809\cdot {\rm ln}^2 E^\star - 2.40411\cdot 10^{-75} \cdot {\rm ln}^{46}E^\star)\right)$} 
                                          &$T_{\rm e}$ & $2.2 \cdot 10^{40}$  & $-4.5$  & 0.0\\
\bottomrule
    \end{tabular*}
    \begin{tablenotes}
\item[{a}] The universal gas constant $\cal R$ must be set to 1.9872	cal/K$\cdot$mol. $A$ has units of $\textrm{cm}^3\cdot(\textrm{mole}\cdot \textrm{s})^{-1}\cdot \textrm{K}^{-n}$. $E$ has units of cal/mole. The rate is given by $A T^n \exp(-E/{\cal R}T).$ The reduced electric field $E^\star$ has units of V~m$^2$. ${\cal A}$ is Avogadro's number and it is approximately $6.022 \cdot 10^{23}~{\rm mol^{-1}}$.
\item[{b}] $\rm M_1=N,~O,~N^+,~O^+$; $\rm M_2=N_2,~O_2,~NO,~N_2^+,~O_2^+,~NO^+$; $\rm M_3= N,~O,~NO,~N^+,~O^+$; $\rm M_4=N_2,O_2,N_2^+,~O_2^+,~NO^+$.
\item[{c}] The forward rates for reactions 10, 11, and 12 are taken from \cite{pf:2007:boyd}. The backward rates for reactions 10 and 11 are sourced from \cite{jgr:2004:sheehan} and \cite{aip:2001:peverall}. The backward  rate for reaction 12 is taken from \cite{aip:1998:peterson} data, using a spline fit listed in \cite{pf:2024:parent}. The backward rates for reactions 24--25 are taken from \cite{nasa:1973:dunn}. Rates for reaction 8 are taken from \cite{jcp:1997:bose} and rates for reaction 9 from \cite{jcp:1996:bose}. The forward rates of reactions 24--25 are found from \cite{jcp:2014:parent}. Other rates are taken from \cite{book:1990:park}. The Townsend ionization rate at very high electric fields $k_{\rm i}=E^\star\cdot(1.875\cdot 10^6 \cdot (-\ln (E^\star)-30)^4+6\cdot 10^{26}\cdot E^\star)\cdot \exp(-7.3\cdot 10^{-19}/E^\star - 5.474\cdot 10^{16}\cdot E^\star)$ in units of cm$^3$/s is obtained from \cite{ps:2005:tarasenko}.
    \end{tablenotes}
   \end{threeparttable}
\end{table*}

\begin{table}[!t]
  \center\fontsizetable
  \begin{threeparttable}
    \tablecaption{Expressions for high-electric-field corrections to electron and ion mobilities.\tnote{a,b}}
    \label{tab:mobilities:Ecorrection}
    \begin{tabular*}{\columnwidth}{@{}c@{\extracolsep{\fill}}c@{}}
    \toprule
    Species & Corrected Mobility, $\rm m^2\cdot V^{-1}\cdot s^{-1}$ \\
    \midrule
    N$_2^+$         & $\min\left( \mu_{\rm N_2^+},~~N^{-1}\cdot 2.03 \cdot 10^{12}\cdot \left(E^\star\right)^{-0.5} \right)$  \alb
    O$_2^+$         &  $\min \left( \mu_{\rm O_2^+},~~N^{-1}\cdot 3.61 \cdot 10^{12}\cdot\left(E^\star\right)^{-0.5}\right)$ \alb
      NO$^+$         & $\min\left( \mu_{\rm NO^+}, N^{-1} \cdot 4.47\cdot 10^{12}\cdot (E^\star)^{-0.5}\right)$\alb
    other ions         & $\min\left( \mu_{\rm i},~~N^{-1}\cdot 0.55\cdot m_{\rm i}^{-0.5} \cdot \left(E^\star\right)^{-0.5} \right)$  \alb
    e$^-$         & \begin{tabular}{@{}l}$(1-\xi)\cdot \mu_{\rm e} + \xi \cdot N^{-1} \cdot \left(4\cdot 10^{19}\cdot(-30 -\ln E^\star)^4+1.3\cdot 10^{40}\cdot E^\star\right) $\\~~~~with~$\xi=\max\left(0,~\min(1,~19+\log_{10}E^\star)\right)$\end{tabular} \\
    \bottomrule
    \end{tabular*}
    \begin{tablenotes}
      \item[a] Notation and units:   $N$ is the total number density of the plasma in 1/m$^3$; $E^\star$ is the reduced effective electric field  ($E^\star \equiv |\vec{E}|/N$) in units of V$\cdot$m$^2$; $\mu_k$ is the uncorrected mobility of species $k$ in $\rm m^2\cdot V^{-1}\cdot s^{-1}$; $m_{\rm i}$ is the ion mass in kg.
      \item[b] The ion mobility correction is taken from \cite{misc:1968:sinnott}. The electron mobility weight factor $\xi$ is such that the corrected mobility is a blend of the uncorrected mobility (function of electron temperature only) and the high-electric-field mobility from \cite{ps:2005:tarasenko} in the range $10^{-19}<E^\star<10^{-18}~ {\rm Vm^2}$. In the range $E^\star>10^{-19}~ {\rm Vm^2}$ the corrected mobility is simply set equal to the high-electric-field mobility.
    \end{tablenotes}
   \end{threeparttable}
\end{table}

\begin{figure}[ht]
    \centering
     \includegraphics[width=0.40\textwidth]{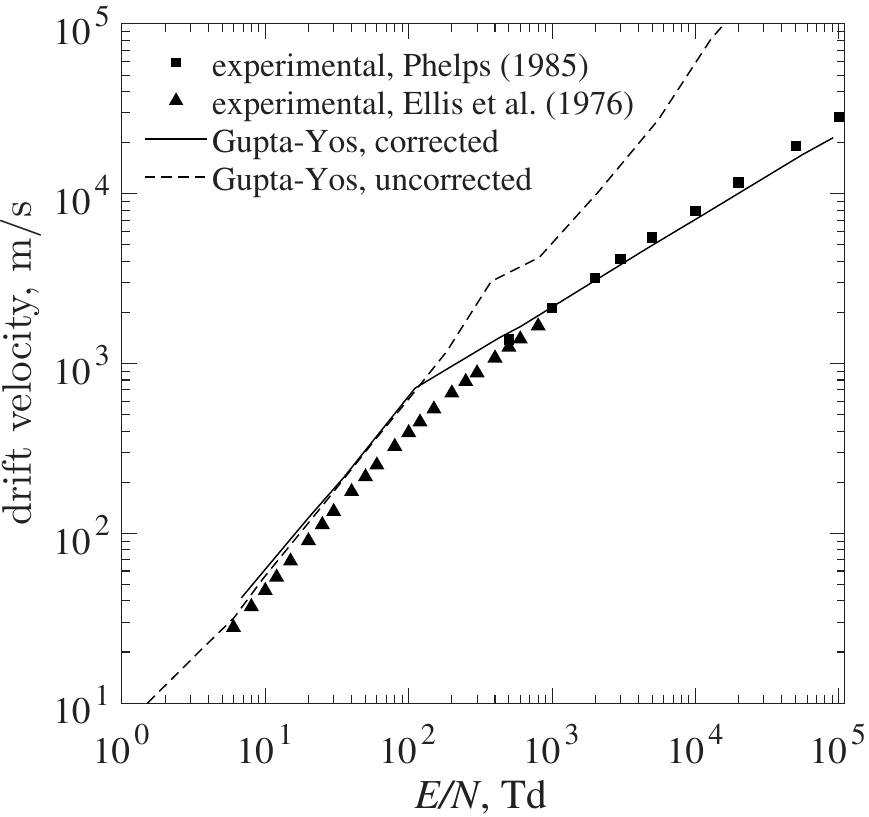}
     \figurecaption{Drift velocity of $\rm N_2^+$ ions in $\rm N_2$ gas at 300 K computed with and without high electric field correction to the ion mobility.}
     \label{fig:drift_N2plus}
\end{figure}

\subsection{High Electric Field Correction to Mobilities}

Electron and ion mobilities are dependent not only on temperature but also on the reduced electric field ($E/N$). While thermal motion drives collisions at low fields, the drift motion becomes dominant when the reduced electric field is high. Consequently, mobility becomes a function primarily of $E/N$ rather than temperature in these regimes. For ions, this transition occurs when ion velocities significantly exceed thermal velocities. In this strong field regime, the ion drift velocity scales as $(E/N)^{0.5}$, with charge-exchange or hard sphere-type cross sections playing a dominant role [\cite{ijms:2009:krylov, misc:1968:sinnott, pr:1954:wannier, jpcr:1991:phelps}]. Experimental data regarding temperature and electric field effects on ion mobility within air plasmas are limited, particularly at reduced electric fields exceeding 2000 Td. Therefore, mobility is often inferred from drift tube experiments measuring ion drift velocity within a specific background gas.

As shown in Fig.~\ref{fig:drift_N2plus}, applying the high-field correction accurately reproduces the data trends from [\cite{jpcr:1991:phelps}] and [\cite{atomicdata:1976:ellis}] for $\rm N_2^+$ ions in nitrogen gas. Notably, the uncorrected Gupta-Yos ion mobility is independent of $E/N$, resulting in a linear dependence of drift velocity on the electric field which fails to capture the physical behavior at high fields. Explicit expressions for these corrections are outlined in Table~\ref{tab:mobilities:Ecorrection}. The corrected mobilities of $\rm N^+$ and $\rm O^+$ are assumed to follow the same mass-dependent expressions as $\rm N_{2}^+$, $\rm O_{2}^+$, and $\rm NO^+$ ions.

\begin{figure}[!t]
    \centering
     \includegraphics[width=0.40\textwidth]{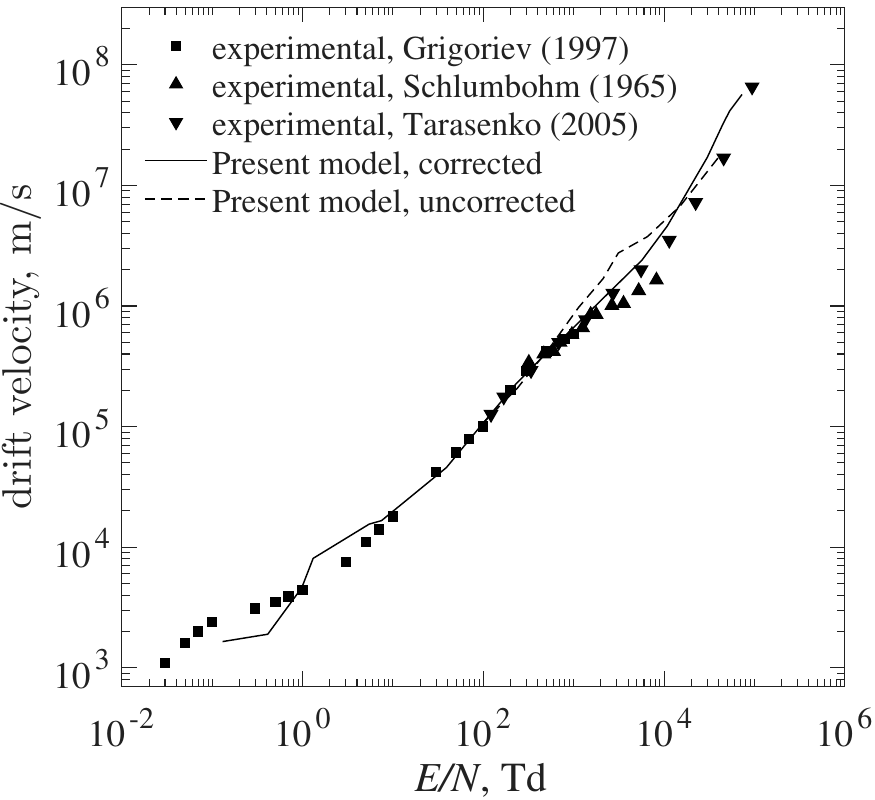}
     \figurecaption{Drift velocity of electrons in $\rm N_2$ gas at 300 K computed with and without high electric field correction to the electron mobility.}
     \label{fig:drift_electron}
\end{figure}

For electron swarms, experimental measurements are reported in terms of the electron drift velocity $u_d$ within a gas mixture:
\begin{equation}u_d = (\mu_{\rm e}N)(T_{\rm e},E/N) \times E/N
\label{eqn:drifteqn}
\end{equation}
where the electron energy or temperature $T_{\rm e}$ is estimated from $E/N$ using the local approximation in a mixture of neutral air species outlined in \cite{pf:2024:parent}. In Fig.~\ref{fig:drift_electron}, we compare electron drift velocities in nitrogen obtained via our physical model with data from \cite{book:1997:grigoriev}, \cite{zfp:1965:schlumbohm}, and \cite{ps:2005:tarasenko}. The application of the high-field correction improves agreement with experimental data, particularly at extremely high discharge conditions exceeding 10,000~Td. In this regime, runaway electrons and near-relativistic conditions significantly affect electron drift velocity (see \cite{ps:2005:tarasenko} for details).

\section{Numerical Methods}

All results presented here were obtained using the in-house CFDWARP (\emph{C}omputational \emph{F}luid \emph{D}ynamics, \emph{WA}ves, \emph{R}eactions, \emph{P}lasmas) code. There are a number of difficulties associated with the integration of the set of equations described in the Physical Model. While the time scale associated with the motion of the neutrals is in the order of microseconds (aerodynamic scale), the time scales associated with chemical reactions and the motion of charged species can be orders of magnitude less. Such disparity is a source of stiffness which is partly overcome by using implicit methods. The block diagonally-dominant alternate-direction-implicit (DDADI) scheme by \cite{aiaaconf:1987:bardina} is found to reduce the number of iterations required to solve the mass, momentum and energy equations for all species. {When including chemical reactions, stability considerations lead to a linearization of the chemical source terms that only incorporate negative Jacobian entries, with the latter being set to zero otherwise. For electron impact ionization reactions function of $E/N$, no linearization is made.}

Another source of stiffness comes from the charge density terms in Gauss' law, specially in quasi-neutral plasma regions where space charge effects are small and results in error amplification (see \cite{book:2022:parent} for a detailed discussion). This can be overcome by obtaining the potential from Ohm’s law instead of Gauss’s law. To ensure that Gauss’s law is satisfied, source terms are added to the ion transport equations. Such a recast of the equations is done without modifying the physical model in any way and is thus simply a convergence acceleration method. {Physically, this equivalence holds because the Ohm-based potential equation is derived directly as a linear combination of the species conservation laws. Consequently, solving the Ohm-based potential equation simultaneously with the species transport equations—augmented with source terms to enforce the Gauss's law constraint—is mathematically equivalent to solving the standard drift-diffusion model coupled with the Poisson equation. This reformulation avoids the error amplification inherent to the Poisson equation in quasi-neutral regions while converging to the exact same physical solution as shown in \cite{book:2022:parent}.} {By solving the neutral flow and charged species transport equations conjunctly in this manner, we can advance the solution using aerodynamic-scale pseudotime steps that are orders of magnitude larger than the physical time scales of electron motion.}

Convergence of the electric field potential equation to steady-state is obtained through a combination of iterative modified approximate factorization (IMAF) by \cite{cf:2001:maccormack} and successive over relaxation (SOR) of \cite{gen:douglas}. {We here found that a combination of 4 IMAF subiterations followed by 100 SOR subiterations yielded good convergence of the electric potential during the pulses. The potential equation is advanced in its own pseudotime step defined in \cite{jcp:2015:parent} by a characteristic length scale (based on the average distance between electrodes) and a reference conductivity.}

{In our time-accurate simulations (during the application of the pulsed electric field), we employ a dual-time stepping approach with the backward Euler implicit formula, performing subiterations in pseudotime until the residuals from both spatial and temporal derivatives are sufficiently minimized. The pseudotime step for the governing equations corresponds to a multiplicative blend of the minimum and maximum CFL conditions across the two dimensions, with exponents of 0.9 and 0.1 respectively. Robust convergence was obtained by starting with a CFL of 0.01 which is increased to 2.0 at a rate of 10\% per iteration.}

The flux discretization employs a hybrid {Harten-Lax-van Leer-Contact} (HLLC) approach based on the HLLC flux by \cite{sw:1994:toro}, reverting to the {Harten-Lax-van Leer} (HLL) flux by \cite{siam:1983:harten} near strong shocks. The blending of HLL and HLLC fluxes is achieved with a shock-stable pressure sensor approach outlined in \cite{cf:2018:simon}. Such hybrid approach avoids the carbuncle issue near strong shocks while minimizing dissipation in viscous layers. The scheme is extended to second-order accuracy through the Van Leer limiter and the Monotonic Upstream-centered Scheme for Conservation Laws (MUSCL) strategy.

{Finally, the overall solution algorithm proceeding within each physical time step can be summarized as follows: (i) the electric potential equation is solved in its own pseudotime loop using 4 IMAF sub-iterations followed by 100 SOR sub-iterations; (ii) the species mass, momentum, and energy conservation equations are solved using the DDADI block-implicit scheme; (iii) the solution is updated in pseudotime, and convergence is checked against the spatial and temporal residuals; and (iv) steps (i)-(iii) are repeated until the residuals drop below the specified tolerance, after which the physical time is advanced.}

\section{Validation Cases}

Here, we cover three hypersonic flow validation cases using the physical model outlined above. These cases correspond to the {Orbital Reentry Experiment} (OREX) flight experiment outlined in \cite{asvpaper:1995:inouye}, the Radio Attenuation Measurement (RAM)  test flight outlined in \cite{nasa:1970:grantham} and \cite{nasa:1972:jones}, and the experimental glow discharge in a hypersonic boundary layer by \cite{thesis:2022:broslawski}.

\subsection{OREX}

 \begin{figure}[!b]
    \centering
     \includegraphics[width=0.45\textwidth]{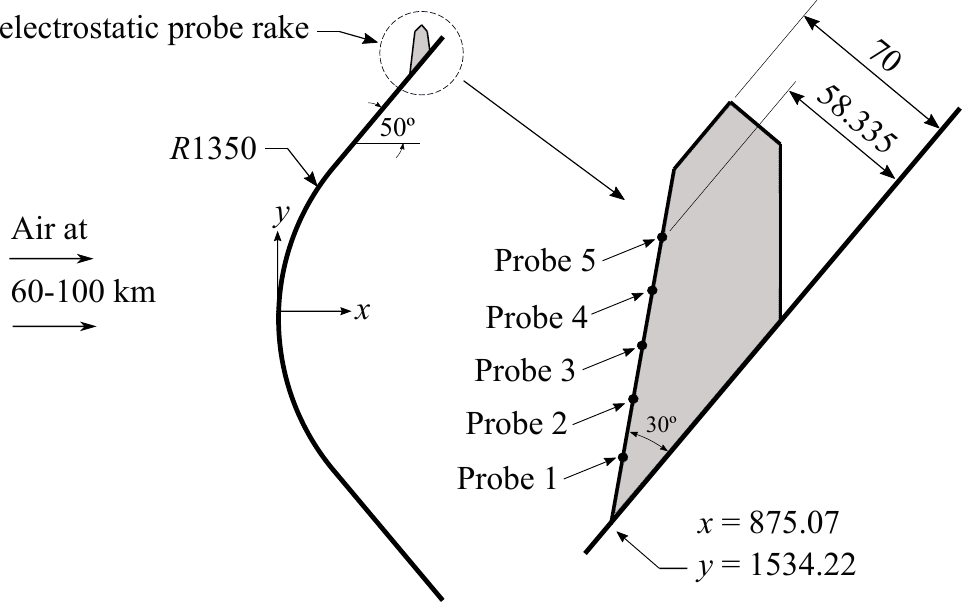}
     \figurecaption{OREX problem setup; dimensions in mm.}
     \label{fig:OREXproblem_setup}
\end{figure}

\begin{table}[!b]
  \center\fontsizetable
  \begin{threeparttable}
    \tablecaption{Freestream and wall boundary conditions for the OREX re-entry trajectory validation cases.}
    \label{tab:OREXconditions}
    \begin{tabular*}{\columnwidth}{c@{\extracolsep{\fill}}cccc}
    \toprule

    altitude, km & $q_\infty$, m/s & $P_\infty$, Pa & $T_{\infty}$, K & $T_{\rm wall}$, K \\
\midrule

   96.7 & 7456.3 & 0.073 & 192.3 & 422 \\
   92.8 & 7454.1 & 0.163 & 188.7 & 492 \\
   88.4 & 7444.3 & 0.231 & 186.9 & 589 \\
   84.0 & 7415.9 & 0.594 & 188.9 & 690 \\
   79.9 & 7360.2 & 1.052 & 198.6 & 808 \\
   75.8 & 7245.7 & 2.172 & 206.8 & 928 \\
   71.7 & 7049.2 & 4.023 & 214.9 & 1078 \\
   67.7 & 6720.3 & 7.892 & 225.9 & 1251 \\
   63.6 & 6223.4 & 14.02 & 237.1 & 1413 \\
   59.6 & 5561.6 & 23.60 & 248.1 & 1519 \\

    \bottomrule
    \end{tabular*}
   \end{threeparttable}
\end{table}

\begin{figure}[ht]
    \centering
     \includegraphics[width=0.35\textwidth]{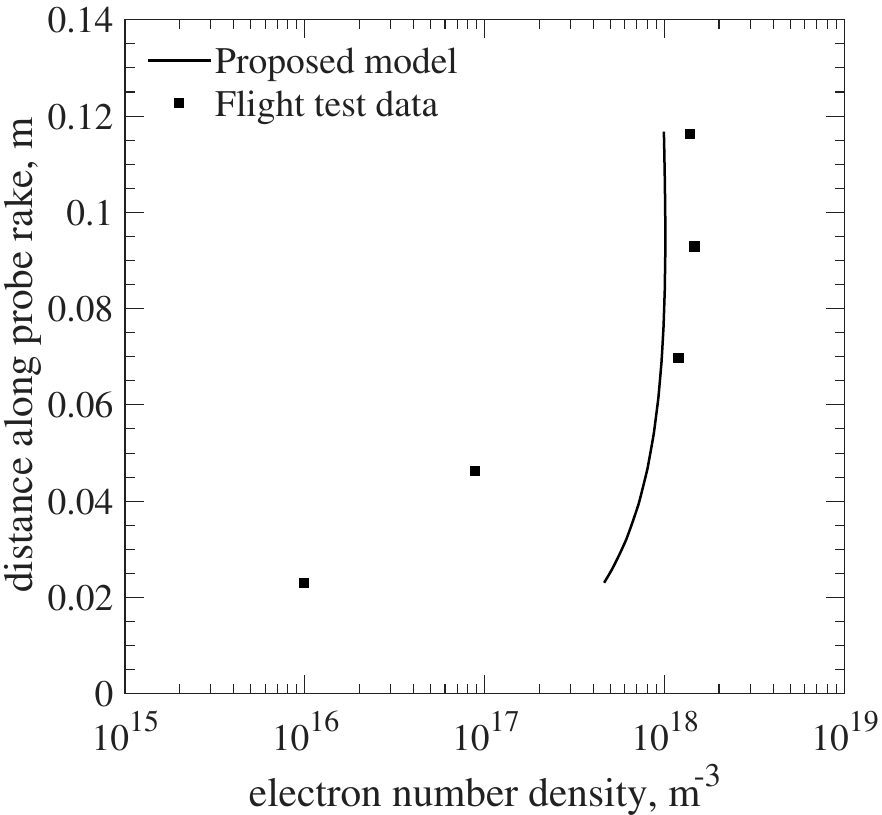}
     \figurecaption{Comparison of computed electron number density with flight measurements along the probe rake at 84 km altitude.}
     \label{fig:OREX_Ne_probe}
\end{figure}

\begin{figure}[ht]
    \centering
     \includegraphics[width=0.37\textwidth]{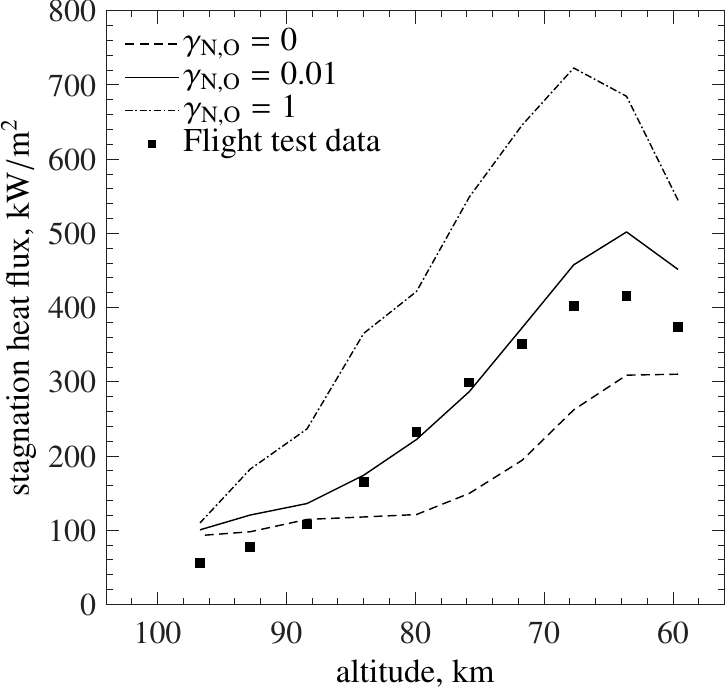}
     \figurecaption{Comparison of stagnation point heat flux between OREX flight data and present CFD simulations.}
     \label{fig:OREX_heatflux}
\end{figure}

The OREX was conducted by the National Aerospace Laboratory and the National Space Development Agency of Japan in 1994 [\cite{asvpaper:1995:inouye, sw:2002:doihara}]. This mission serves as an excellent validation case as it is rich with plasma density and heat flux data across several altitudes in strong thermal non-equilibrium. The OREX vehicle was equipped with five electrostatic probes to measure saturated ion currents and infer electron density upon return. A schematic of the OREX geometry and electrostatic rake is shown in Fig.~\ref{fig:OREXproblem_setup}, while Table~\ref{tab:OREXconditions} outlines the freestream properties and wall temperatures for the altitude range covered in this validation.

A comparison of electron density computed along the probe rake location at an altitude of 84 km is shown in Fig.~\ref{fig:OREX_Ne_probe} and compared to the inferred electron number density in flight. We find good agreement at probing locations away from the OREX surface. {The observed discrepancy between flight test data and numerical results at probes 1 and 2 (closest to the wall) is consistent with the findings of \cite{aiaapaper:1996:gupta} and \cite{sw:2002:doihara}, who also reported a similar mismatch in this region without identifying a definitive cause. We attribute this to the limitations of the experimental diagnostics closer to the surface, where the proximity to the vehicle body likely introduces deviations from the idealized collection model assumptions.  This leads to large uncertainties when inferring electron density from the measured voltage and current in this region.}

In-flight heat flux measurements were obtained by monitoring the temperature of the surface at the stagnation point as a function of time. In Fig.~\ref{fig:OREX_heatflux}, we compare numerical results with OREX flight test data for stagnation point heat flux over an altitude range of 60 km to 97 km. The results highlight the significant role played by the catalytic recombination coefficients of atomic nitrogen and oxygen. Notably, the comparison reveals a varying trend: we observe better agreement with a low surface catalysis model at higher altitudes, transitioning to better agreement with a high surface catalysis of 0.01 at lower altitudes. This behavior is physically consistent, as surface catalycity is well known to increase as a function of time and temperature during the reentry process. Further, such is in accordance with previous post-flight analyses such as the one by \cite{htmp:2005:enzian} which report an effective catalytic recombination coefficient on the order of 0.01 for the C/C composite coating used in the OREX heat shield.

\begin{figure}[!b]
    \centering
    \includegraphics[width=0.35\textwidth]{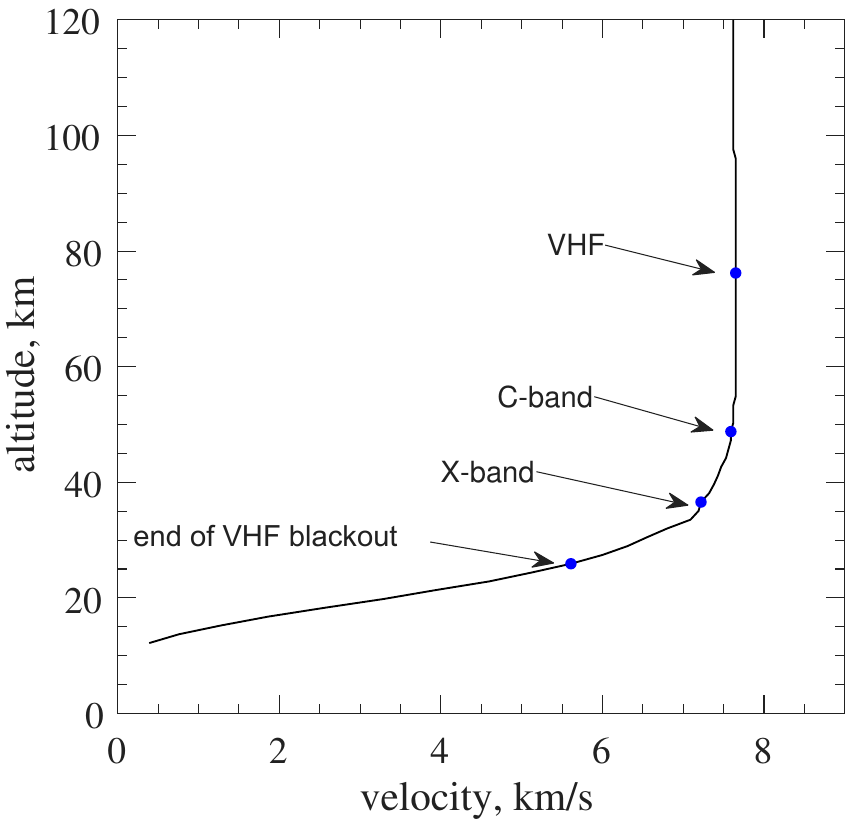}
     \figurecaption{RAM-C-II flight envelope with communication blackout onset.}
     \label{fig:RAMCII_FLIGHTENVELOPE}
\end{figure}
\begin{figure}[!b]
    \centering
     \includegraphics[width=0.35\textwidth]{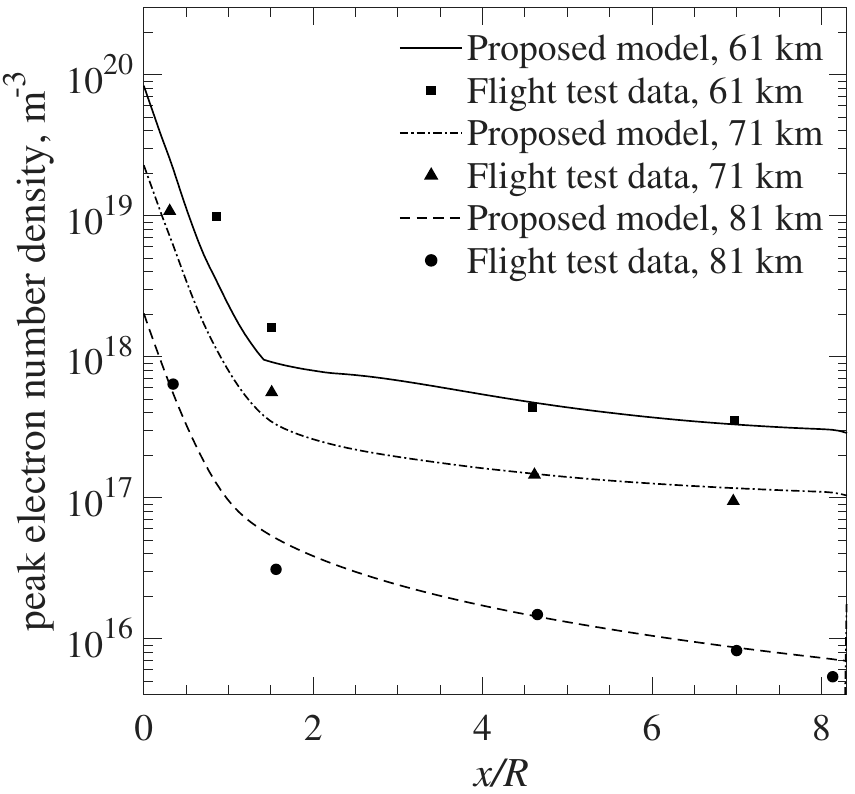}
     \figurecaption{Axial distribution of peak electron number density for RAM-C-II at various altitudes; Distance $x$ is normalized by the nose cap radius $R$.}
     \label{fig:RAMCII_Ne}
\end{figure}

\subsection{RAM-C-II}

The 1960s NASA Radio Attenuation Measurement (RAM) experiments were conducted to investigate reentry plasma attenuation and radio-frequency blackout. The RAM-C-II vehicle has a blunt-wedge shaped geometry with a nose radius of 0.1524 meters, a half-cone angle of 9 degrees,
and a body-length 8.5 times the nose radius. Microwave reflectometers positioned along the body were used to measure the maximum
number density at given streamwise stations. Flight-test data from two altitudes of 61, 71 and 81 km are compared here with numerical results. In Fig.~\ref{fig:RAMCII_FLIGHTENVELOPE} we show the RAM-C-II flight envelope with indicated communication blackout onset adapted from \cite{nasa:1970:grantham}. In the altitude range for which there is experimental electron density data (60-80 km) the velocity is constant and equal to about 25,000 ft/s or 7.62 km/s.

In Fig.~\ref{fig:RAMCII_Ne} we show a comparison of numerical results with flight test data on the basis of the maximum electron number density along the RAM-C-II axis, normalized with the nose cap radius such that $x/R=0$ corresponds to the stagnation point. Overall excellent agreement is found at altitudes of 61 and 71 kilometers, in particular, for the peak electron density along the wedge region. At 81 km, the use of the forward control temperature $\sqrt{TT_{\rm v}}$ recommended in \cite{jtht:1993:park} for the dissociation reactions leads to very good agreement despite the very low density conditions.

\begin{figure}[!b]
    \centering
     \includegraphics[width=0.45\textwidth]{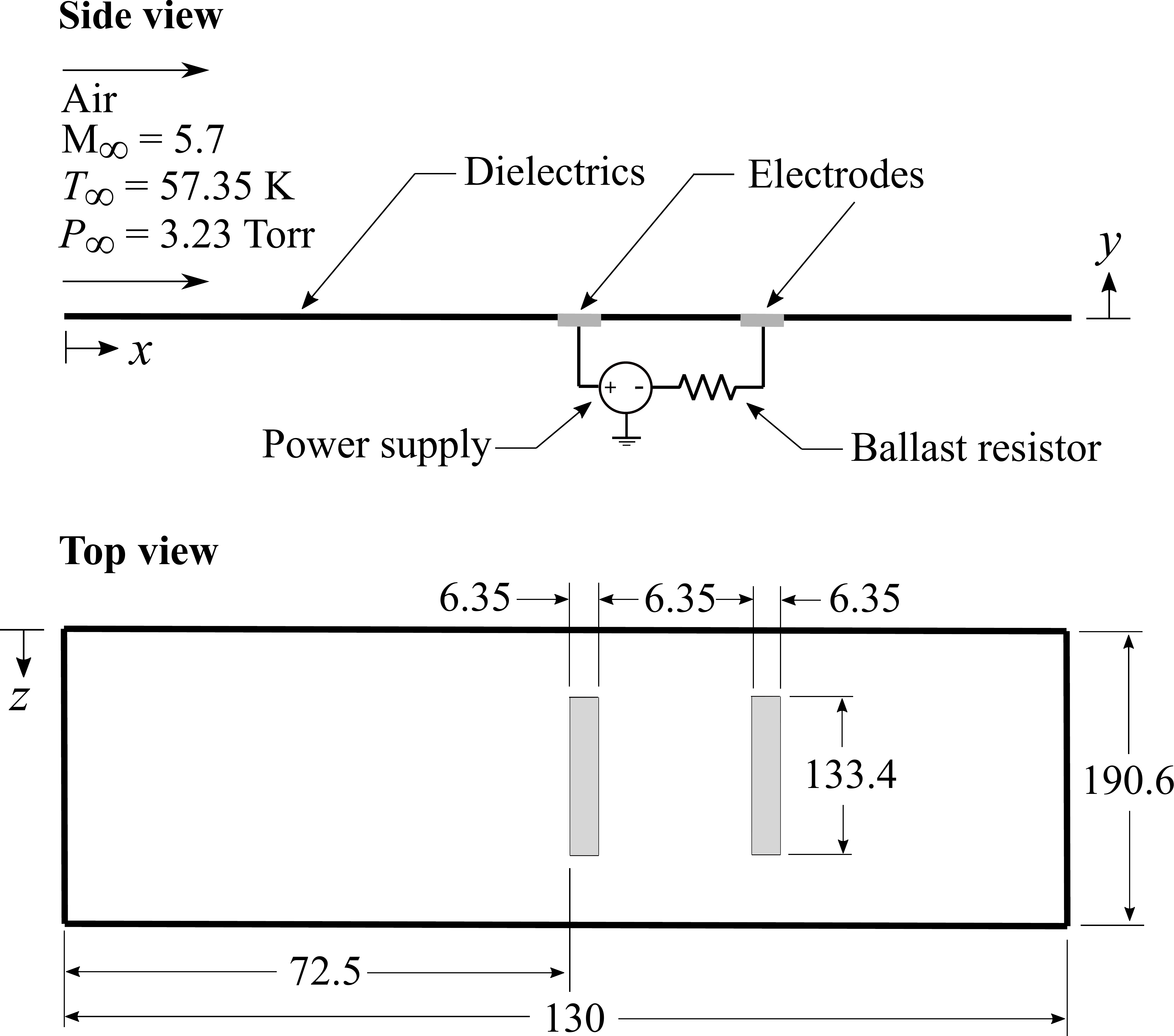}
     \figurecaption{Problem setup for the experimental glow discharge validation case; all dimensions in mm. Reproduced from Phys.\ of Fluids 36, 8 (2024), with the permission of AIP Publishing.}
     \label{fig:discharge_setup}
\end{figure}
\begin{figure}[!t]
    \centering
     \includegraphics[width=0.35\textwidth]{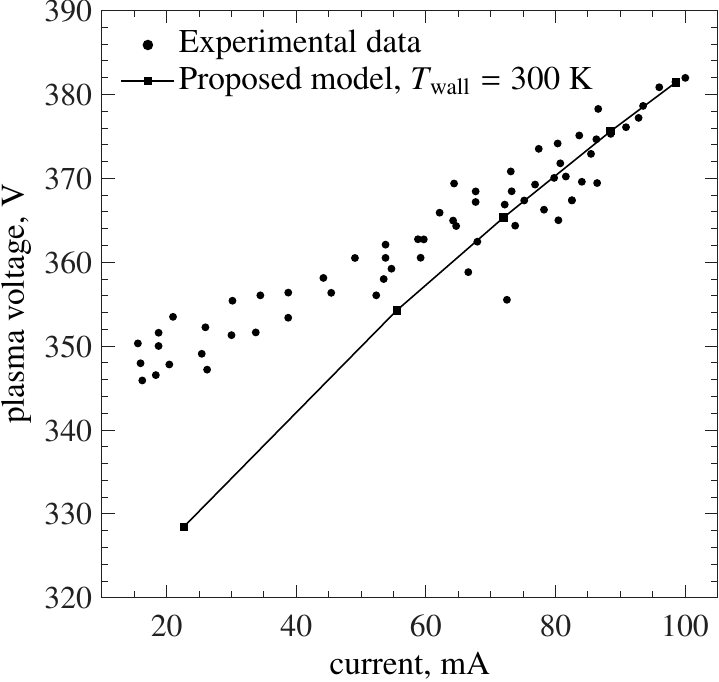}
     \figurecaption{Comparison of plasma voltage--current characteristic with experimental data for the hypersonic boundary layer glow discharge case; with the wall temperature fixed to 300 K.}
     \label{fig:broslawski_voltagecurent}
\end{figure}

\subsection{Glow Discharge in Hypersonic Boundary Layer}

The third and last validation case involves a direct-current glow discharge acting on a hypersonic boundary layer, investigated experimentally by \cite{thesis:2022:broslawski}. The problem setup and flow conditions are shown in Fig.~\ref{fig:discharge_setup}. In the experiments, the wall temperature was seen to vary between 300~K and 450~K along the plate and also between experimental runs. Because the exact temperature distribution is not known and for simplicity, we here fix the wall temperature to 300~K. This problem is simulated in 2D because the thickness of the plasma sheath and boundary layer is orders of magnitude less than the electrode depth, and edge effects near the center of the domain are expected to be negligible.

With a ballast resistance set to 30~k$\Omega$ in the experiment, most of power is lost in the resistance rather than the plasma. Consequently, the plasma voltage, or the voltage difference across electrodes, is highly sensitive to the physical model. By varying the power supply voltage from 1000 to 3300~V, we obtain the plasma voltage--current characteristic and compare them with experimental results.

In Fig.~\ref{fig:broslawski_voltagecurent} we show a comparison of experimentally measured plasma voltage with numerical results. Here, we fixed the secondary electron coefficient at the electrode to a value of 0.45. Such is within range of experimental data of effective electron yield per ion \cite[Fig.~5]{psst:1999:phelps} for a similar reduced electric field (approximately 20,000~Td) at a high--current cathode. Good agreement is found in the voltage--current characteristic at high current. {The discrepancy observed at lower currents is primarily attributed to the limitations of the Local Field Approximation (LFA). While  \cite{pf:2024:parent} demonstrated that combining the Local Energy Approximation (LEA) with a variable wall temperature (300--450~K) yields superior agreement across the full range, varying the wall temperature in the LFA model did not yield similar improvements in the low-current regime. We nonetheless retain the LFA to maintain consistency with the physical model used for the primary results of this study, as the LEA encounters stability constraints at the high reduced electric fields characteristic of the reentry cases.}

\section{Problem Setup}

\begin{figure}[!t]
    \centering
     ~\\~\\~\\
     \includegraphics[width=0.47\textwidth]{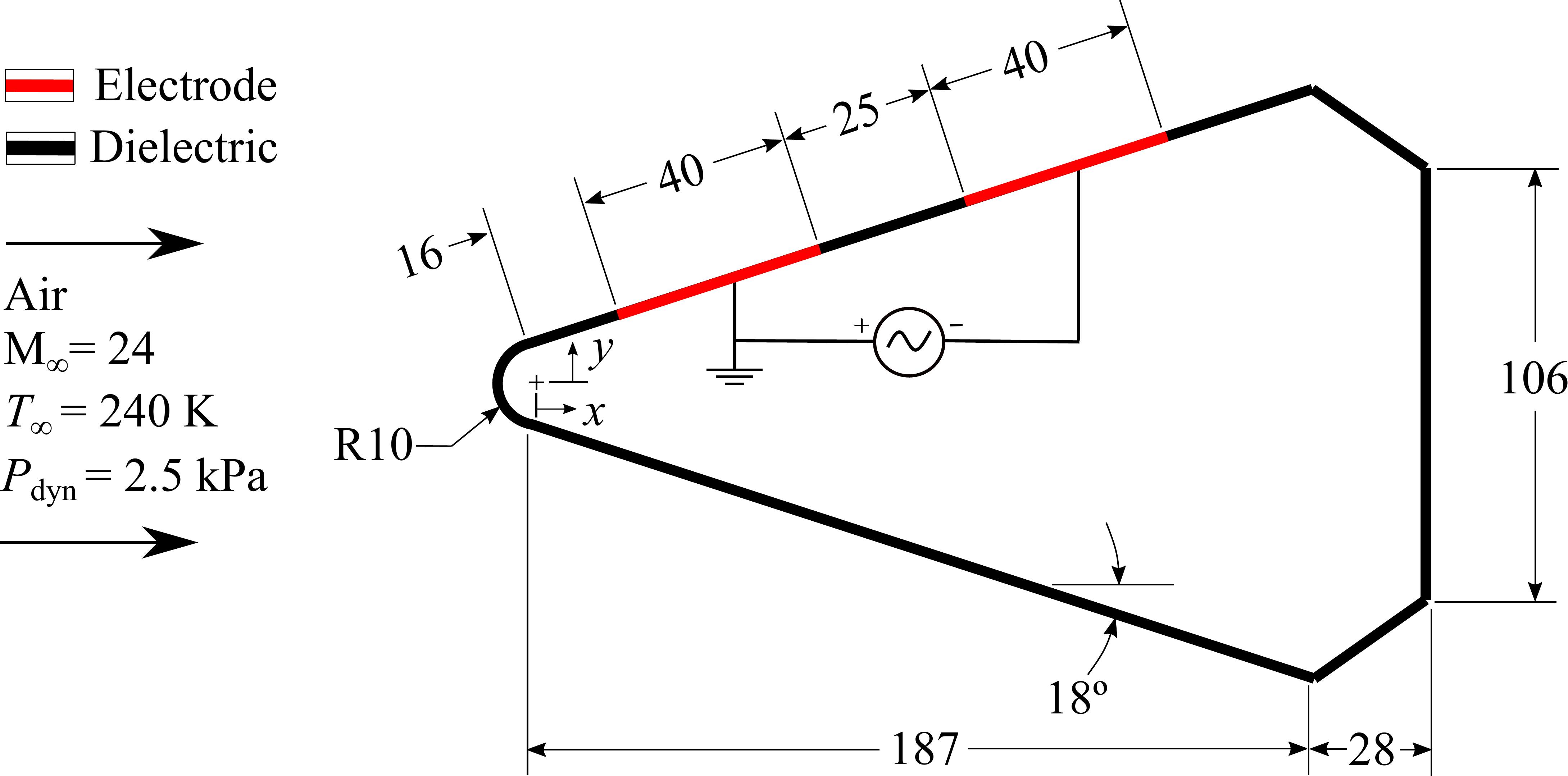}
     \figurecaption{Problem setup for the sharp leading edge waverider under consideration; dimensions in mm.}
     \label{fig:problem_setup}
\end{figure}

\begin{figure}[!t]
    \centering
     ~\\~\\
     \includegraphics[width=0.35\textwidth]{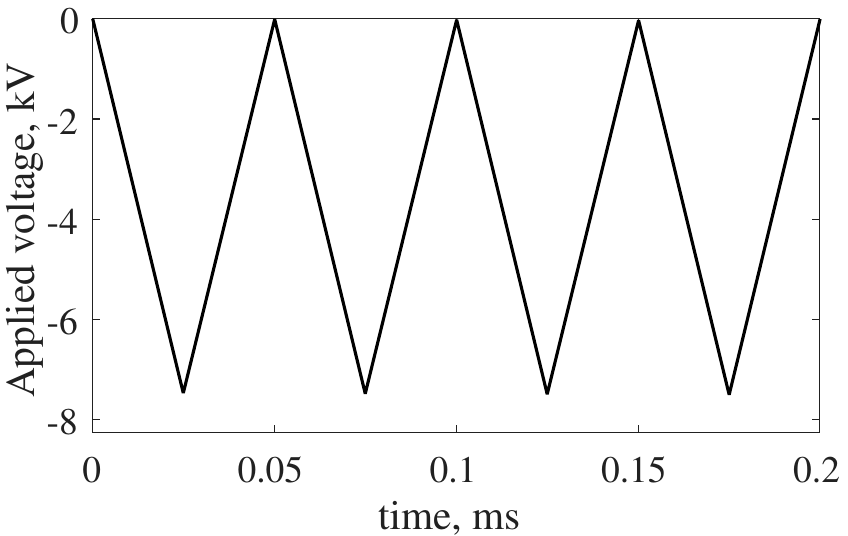}
     \figurecaption{Baseline pulsed voltage cycles applied to the active electrode.}
     \label{fig:applied_volt}
\end{figure}

The problem tackled herein corresponds to the hypersonic flow interacting with a two-dimensional, 18$^\circ$ wedge with a round leading edge as shown in Fig.~\ref{fig:problem_setup}. The Mach number is 24, the dynamic pressure is set to 2.5~kPa and the freestream temperature is of 240~K. This corresponds to an altitude of about 68 kilometers and a reentry velocity of 7.45~km/s. Unless otherwise specified, the wall temperature is set here to 1400~K. Two electrodes of 40~mm in length are placed on the wedge, with the first electrode grounded and another active electrode placed after the latter and separated by 25~mm.

First, a steady-state solution of the plasma flow surrounding the hypersonic vehicle is obtained with both electrodes grounded. Starting from the latter solution, time-accurate simulations are run by applying the pulsed voltage shown in Fig.~\ref{fig:applied_volt} to the active downstream electrode immersed in the plasma, with 0.2~ms of simulation time covering a total of 4 pulse cycles.

\begin{figure}[!t]
     \centering
     ~\\~\\
     \subfigure[]{\includegraphics[width=0.36\textwidth]{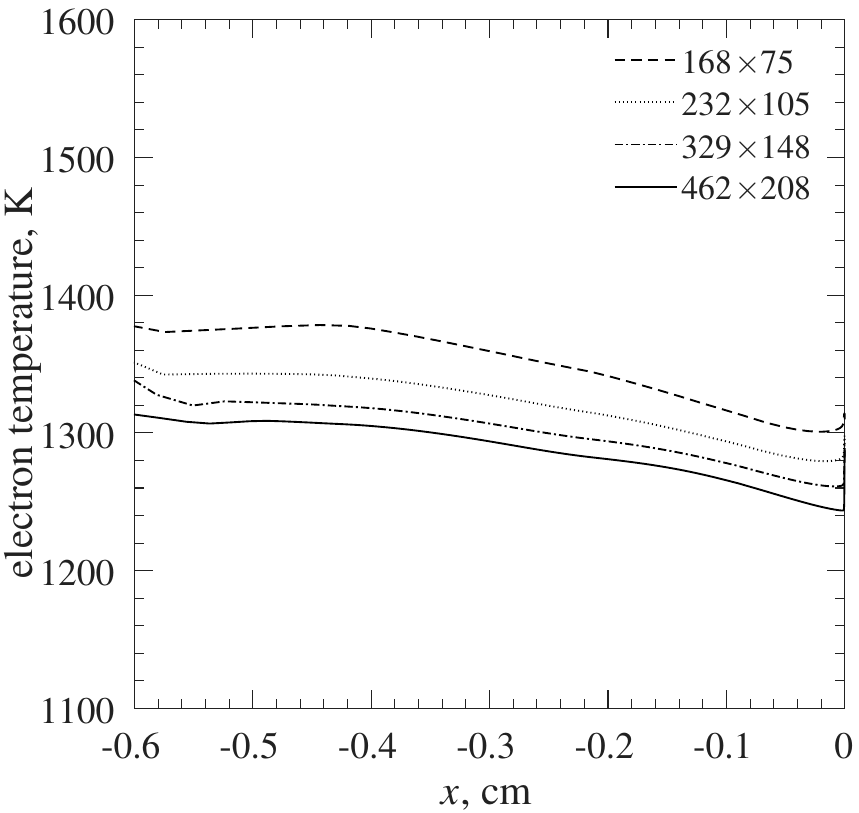}}
     \subfigure[]{\includegraphics[width=0.36\textwidth]{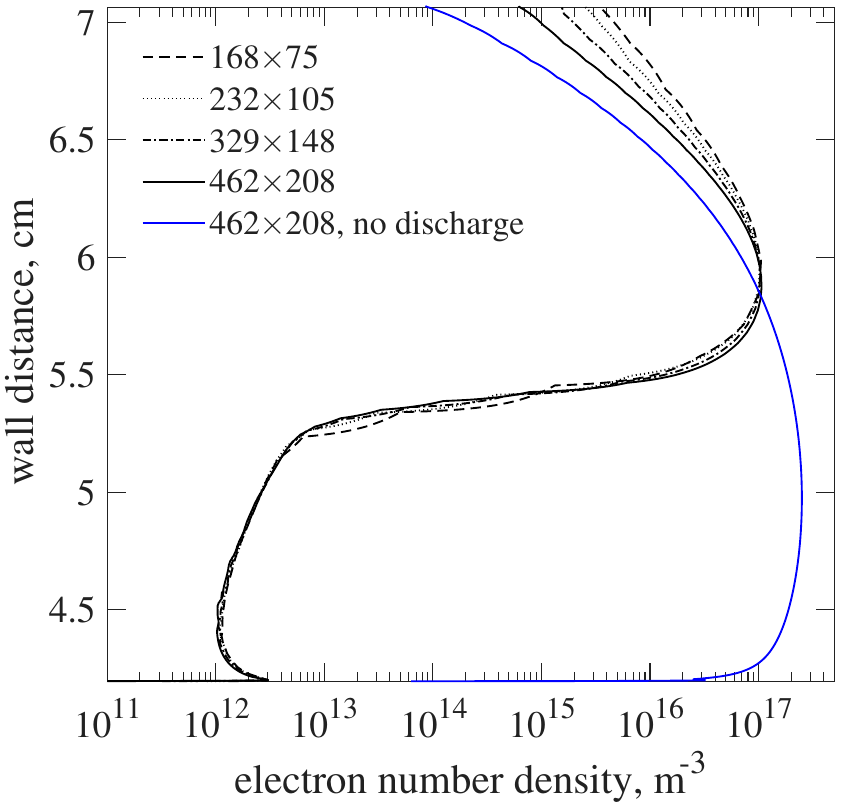}}
     \figurecaption{Effect of grid size on (a) electron temperature along the stagnation line and (b) electron number density profiles extracted along the wall-normal at the midpoint of the active electrode under peak pulse voltage (7.5~kV).}
     \label{fig:numerical_error_fig}
\end{figure}

\section{Numerical Error Assessment}

To determine the grid resolution required to minimize numerical error, we performed a convergence study starting with a baseline grid of $168\times 75$ and systematically increasing the number of grid lines by a factor of 1.41 in both directions. Figure~\ref{fig:numerical_error_fig} presents the electron temperature along the stagnation line and electron density profiles extracted from the center of the cathode during the middle of the pulse, corresponding to the moment of peak electric field.

Electron temperature proved to be the parameter most sensitive to grid resolution. By applying the Grid Convergence Index (GCI) method from \cite{aiaa:1998:roache}, we determined an observed order of accuracy of 1.5, which is consistent across the four meshes and indicates the solution is within the asymptotic range of convergence. Based on this analysis, the error associated with $T_{\rm e}$ on the finest mesh ($462\times208$) is estimated to be approximately 4\%. {The temperature variations of approximately 100~K observed between grid levels in Fig.~13a correspond to the reduction of discretization error as the mesh is refined. These variations diminish monotonically with increasing grid density, confirming that the solution is converging toward a grid-independent state.}

In contrast, the thickness of the non-neutral cathode sheath demonstrates little sensitivity to grid resolution. Even the coarsest mesh considered predicts a sheath thickness within 3\% of the finest mesh. This robustness is significant because the sheath thickness is the governing parameter that defines the extent of the region where electron density drops by orders of magnitude compared to the bulk plasma. This behavior is not unique to this specific problem; we observed a similar trend when simulating the Broslawski test case in Section III-C.








\begin{figure*}[!t]
     \subfigure[Initial steady-state flowfield]{\includegraphics[width=0.49\textwidth]{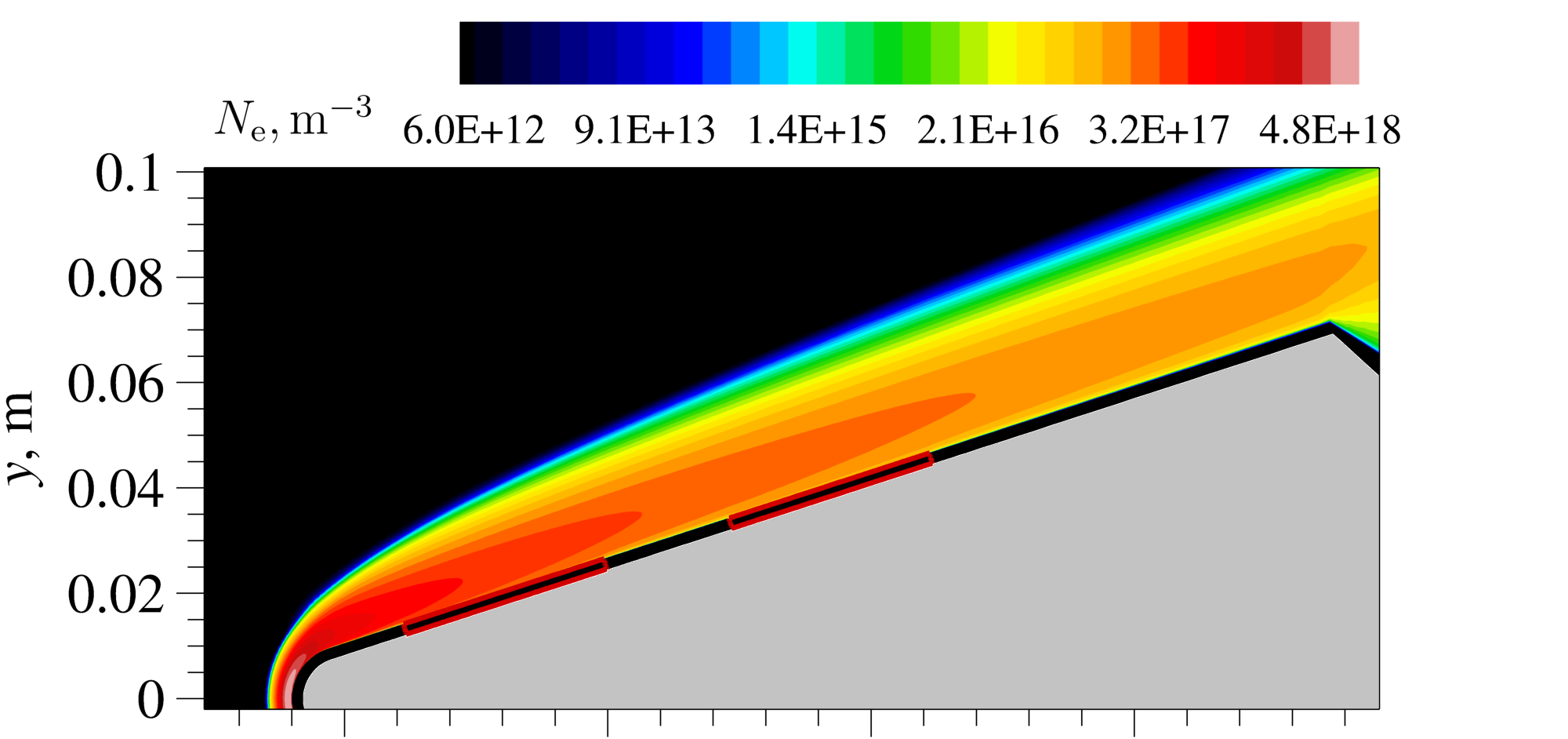}}
     \subfigure[Baseline discharge at peak pulse voltage (7.5 kV)]{\includegraphics[width=0.49\textwidth]{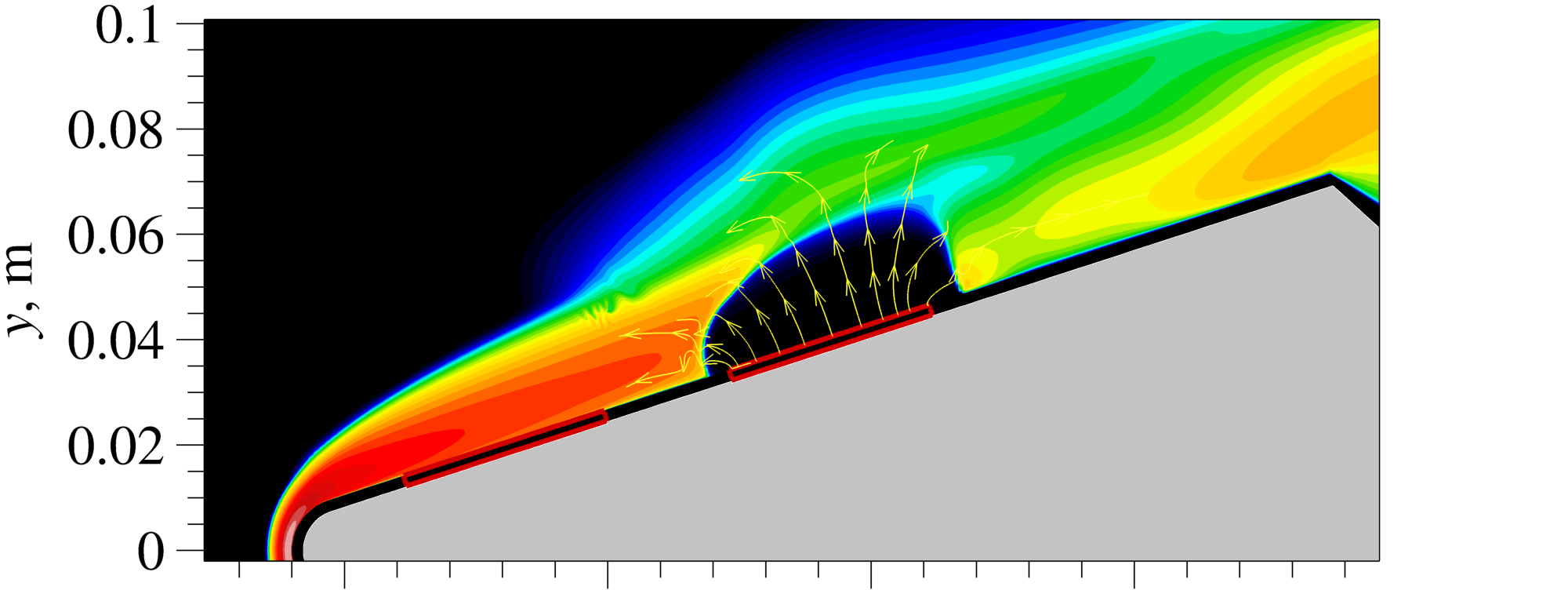}}
     \subfigure[Discharge with correction to ion mobilities]{\includegraphics[width=0.49\textwidth]{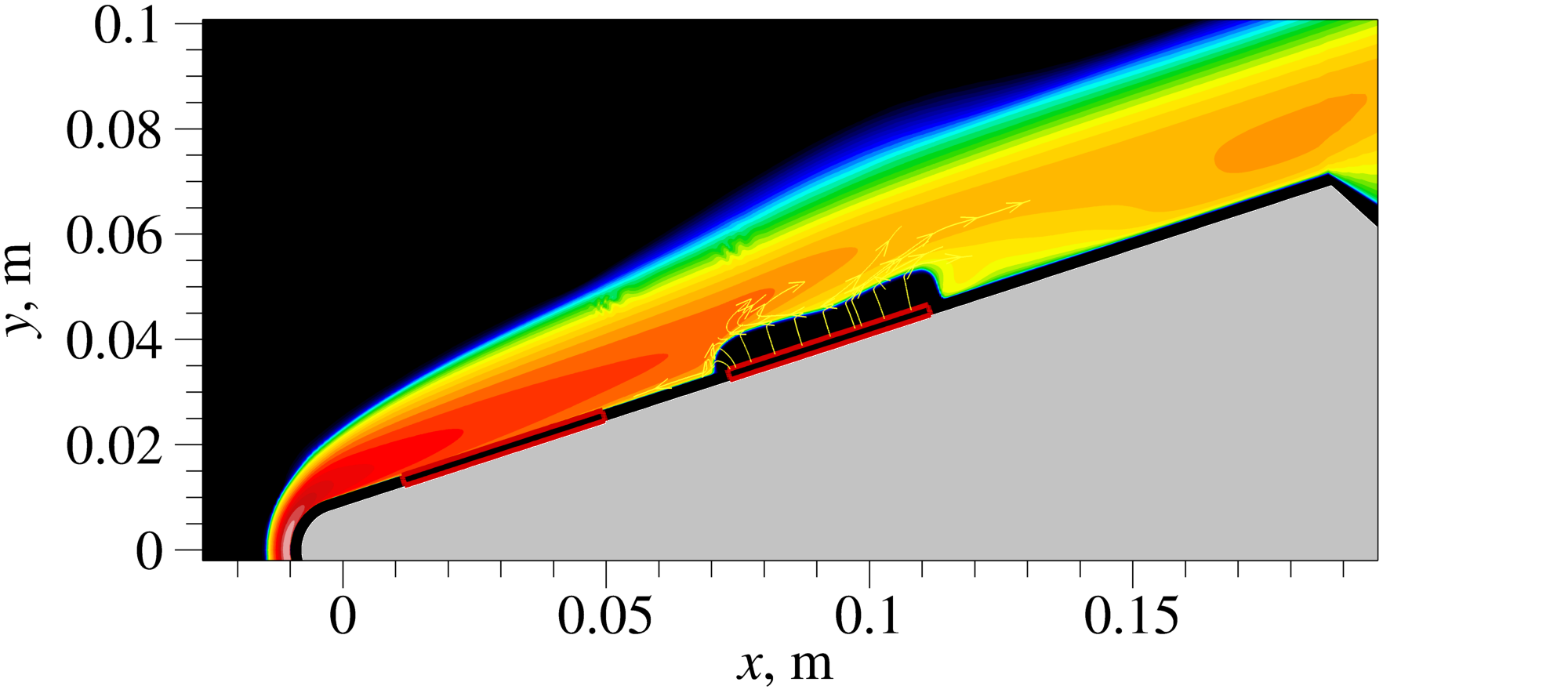}}
    ~~~\subfigure[Discharge with uncorrected electron mobilities]{\includegraphics[width=0.49\textwidth]{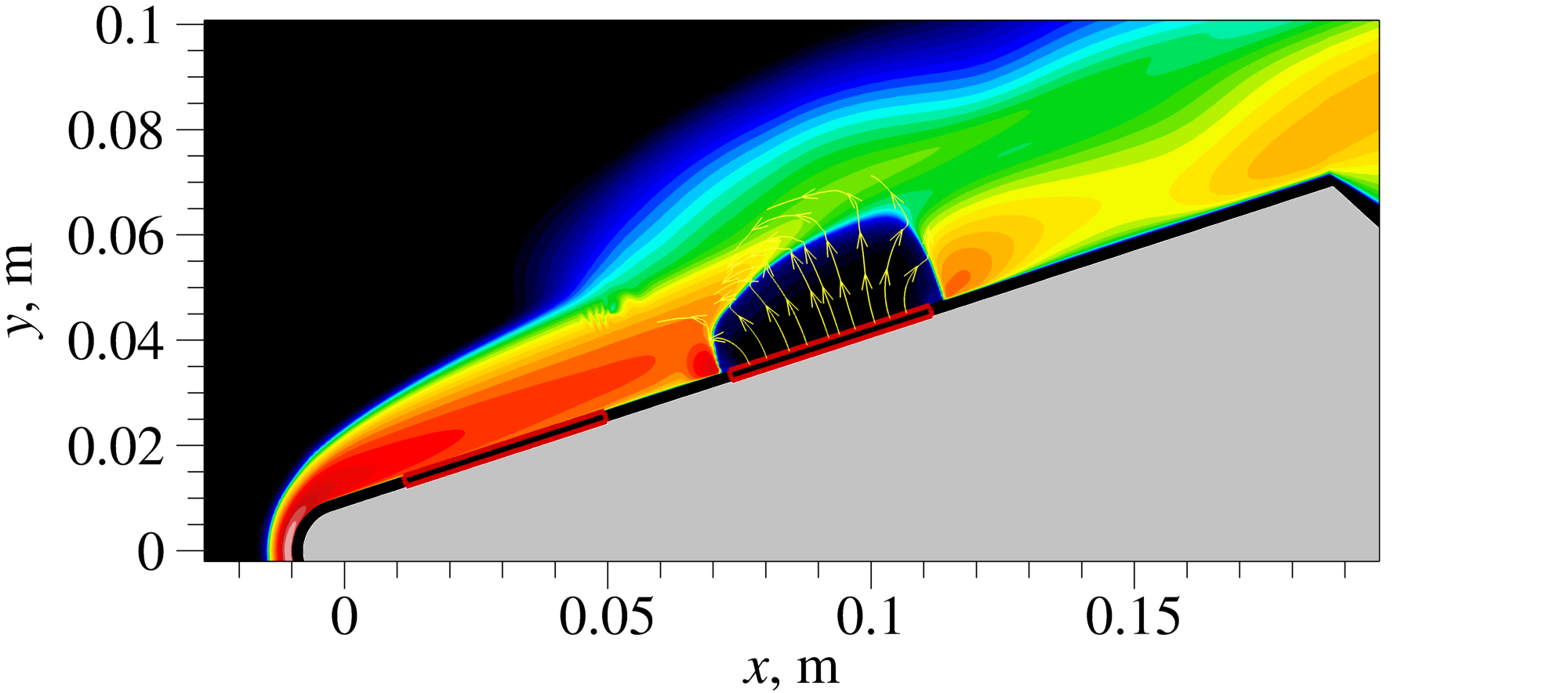}}
     \figurecaption{Electron number density contours: (a) Initial steady-state flowfield; (b) Baseline discharge at peak pulse voltage (7.5 kV); (c) Discharge with high-$E/N$ correction to ion mobilities; (d) Discharge with uncorrected electron mobilities. Electron velocity streamlines are superimposed in the cathode region.}
     \label{fig:Ne_contours_Efield_effects_PM}
\end{figure*}

\section{Performance {Metrics}}
\label{sec:performance_metrics}

In our results, we will seek to measure the attenuation of telemetry signals as they penetrate the plasma layer left after applying a pulsed electric field. Because the properties are not uniform across the thickness of the hypersonic plasma, it is difficult to measure the attenuation exactly. The attenuation is not only dependent on the local plasma frequency but also on the thickness of the plasma layer itself. Rather, we can estimate the signal attenuation by assuming that the hypersonic plasma thickness on top of the cathode can be approximated as an uniform slab of plasma with average properties measured from our solution. 

{To facilitate this analysis, we first introduce operational definitions for the distinct plasma regions observed in the flowfield. The boundary between the non-neutral sheath and the quasi-neutral plasma is defined as the largest distance from the surface where the local net charge density $\rho_{\rm c}$ exceeds 1\% of the peak electron density found in the adjacent bulk (quasi-neutral) plasma layer. Based on this threshold, we define the ``sheath thickness'' $d_{\rm s}$ as the normal distance from the cathode surface to this boundary. Consequently, the ``plasma thickness'' $d_{\rm p}$ is defined as the remaining distance extending from the sheath boundary to the shock front.}

In this case, the relationship between the signal intensity $I_0$ and the intensity that reaches the vehicle surface $I$ (transmitted) has the following expression taken from \cite{cpc:2006:jin} for a non-magnetized uniform plasma:
\begin{equation}
   \log \frac{I}{I_0}
= 
-\,2 d_{\rm p}\,\frac{\omega_{\rm s}}{c}
\sqrt{
\frac{
\sqrt{
\left(
1 - \frac{\omega_{\rm p}^4}{\omega_{\rm s}^4 + \nu_{\rm en}^2 \omega_{\rm s}^2}
\right)^{\!2}
+ 
\left(
\frac{\nu_{\rm en}\,\omega_{\rm p}^2\,\omega_{\rm s}}{\omega_{\rm s}^4 + \nu_{\rm en}^2 \omega_{\rm s}^2}
\right)^{\!2}
}
-\!
\left(
1 - \frac{\omega_{\rm p}^4}{\omega_{\rm s}^4 + \nu_{\rm en}^2 \omega_{\rm s}^2}
\right)
}{2}
}
\label{eqn:atten}
\end{equation}
where $I/I_0$ is the ratio of transmitted and incident signal intensity, $d_{\rm p}$ is the plasma layer thickness {defined above}, $\omega_{\rm p}$ is the circular plasma frequency, $\omega_{\rm s}$ is the signal circular frequency and $\nu_{\rm en}$ is the electron-neutral collision frequency. The collision frequency is obtained from the average electron number density and electron mobility in this region to evaluate Eq. (\ref{eqn:atten}). 

Another parameter that we will use is the average power deposition to the plasma due to the pulsed electric field divided by the cathode surface area. This is computed as follows:
\begin{equation}
{\cal P} \equiv \displaystyle\dfrac{1}{{\cal T} S}\int_{t=0}^{\cal T}\int_{\Vol} \vec{E}\cdot \vec{J}~{\rm d}\Vol{\rm d}t
    \label{eqn:power_deposited}
\end{equation}
with $\cal P$ the power per cathode area, $\cal T$ the period of one cycle (0.05 ms), $\Vol$ the volume of the computational domain, and $S$ is the surface area of the cathode touching the plasma.

\begin{figure}[!h]
     \centering
     \subfigure[Baseline discharge at peak pulse voltage]{\includegraphics[width=0.43\textwidth]{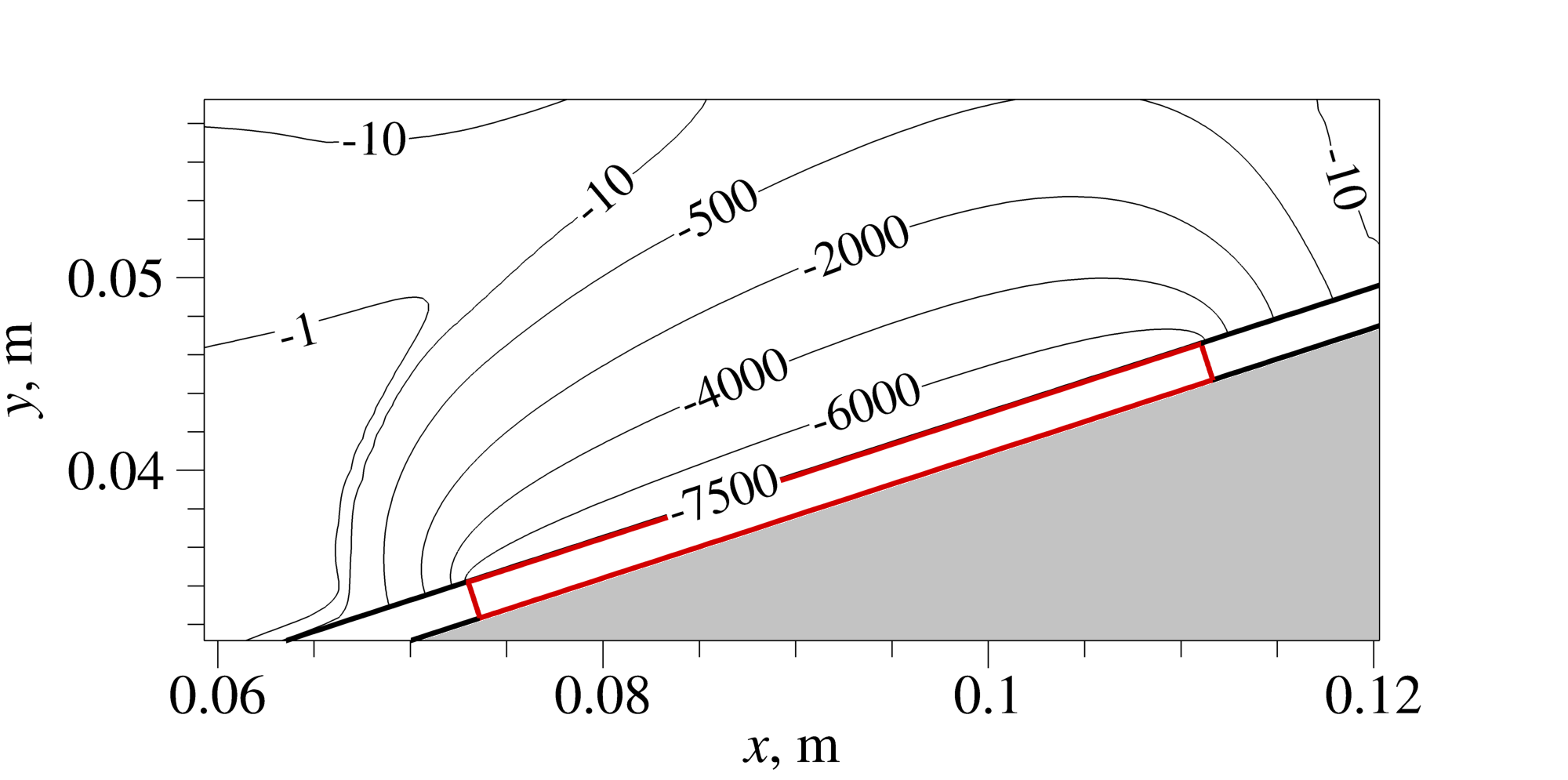}}
     \subfigure[Discharge with correction to ion mobilities]{\includegraphics[width=0.43\textwidth]{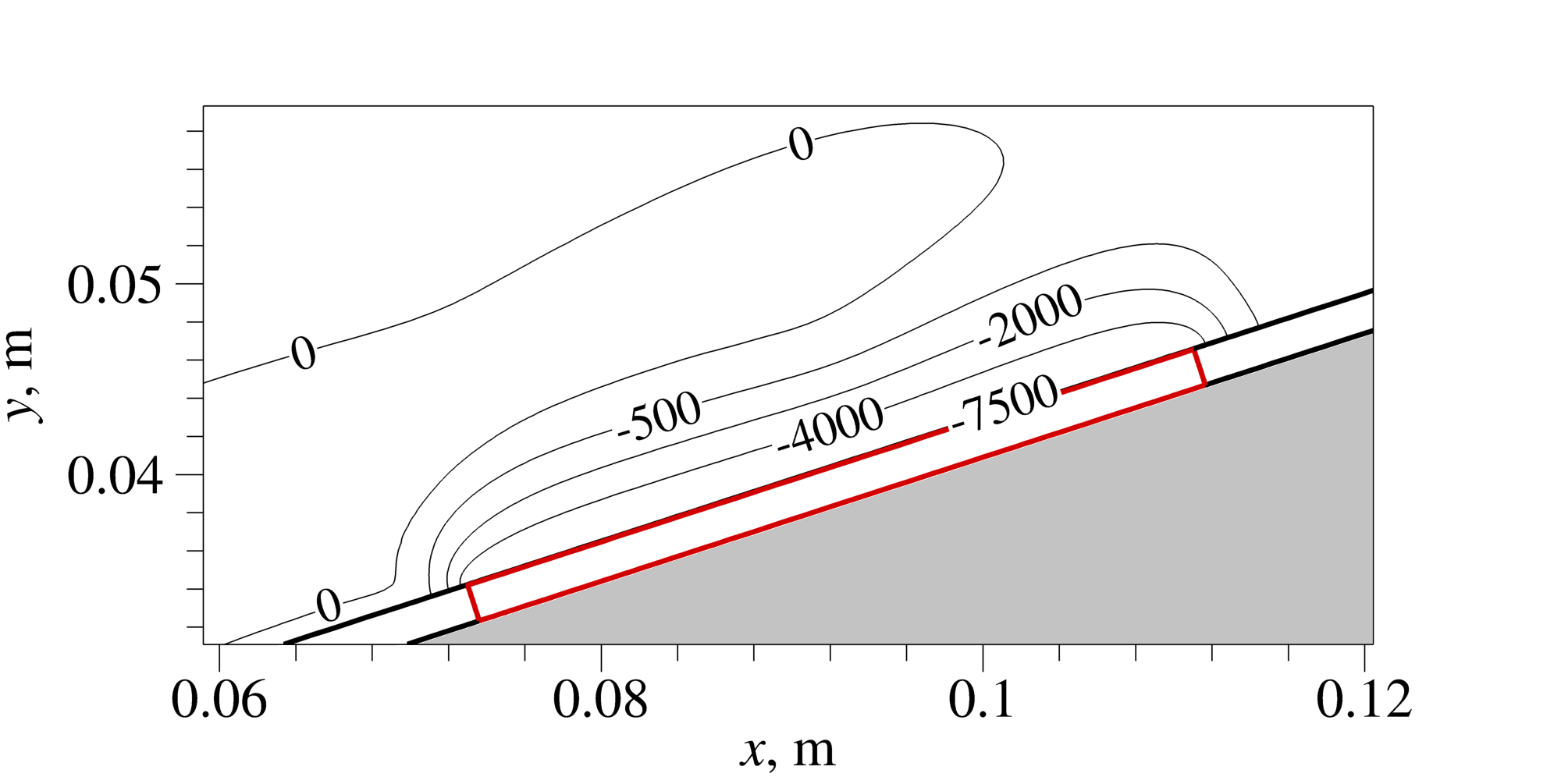}}
     \figurecaption{Electric potential $\phi$ (V) contours near the cathode: (a) Baseline discharge at peak pulse voltage; (b) Discharge with correction to ion mobilities.}
     \label{fig:cathode_drop}
\end{figure}

\section{Results and Discussion}

The following results detail the interaction between high-voltage pulses and the Mach 24 flowfield around a re-entry vehicle at 68 km altitude. As this represents a foundational study of these fully-coupled interactions, the flight Mach number, altitude, and triangle-waveform pulse cycles are held constant to provide a consistent framework for comparison. We investigate the sensitivity of the electron depletion region to three primary factors: ion mobility corrections, electron mobility corrections, and the secondary electron emission (SEE) coefficient. Our analysis is centered on how these physical parameters alter the mitigation performance, measured through signal attenuation and the requisite power deposition. Individual parameters are varied systematically from a baseline configuration—defined by a 7.5~kV peak and corrected electron mobility—to isolate their respective roles in the discharge physics.

\begin{table*}[!t]
  \center
  \begin{threeparttable}
\tablecaption{Estimated signal attenuation through the plasma layer for various frequency bands, with and without pulsed electric field application.\tnote{a,b}}
    \label{tab:atten}
    \fontsizetable
    \begin{tabular*}{\textwidth}{l@{\extracolsep{\fill}}cccccccccc}
    \toprule
~&~&~&\multicolumn{4}{c}{uncorrected ion mobility} & \multicolumn{4}{c}{corrected ion mobility} \\
 \cmidrule(lr){4-7}\cmidrule(lr){8-11}
band & \multicolumn{1}{c}{$f_{\rm s}$, GHz} & applied voltage, kV & $d_{\rm p}$, cm & $N_{\rm p},~{\rm m^{-3}}$ & $f_{\rm p}$, GHz & $I/I_0$ & $d_{\rm p}$, cm & $N_{\rm p},~{\rm m^{-3}}$ & $f_{\rm p}$, GHz & $I/I_0$ \\ 
\midrule

VHF& 0.10   & -0.0 & $2.53$  & $2.31\cdot10^{17}$ & $4.33$  & $0.01$0 & $2.53$  & $2.31\cdot10^{17}$ & $4.33$  & $0.010$ \\

UHF& 0.50   & -0.0 & $2.53$  & $2.31\cdot10^{17}$ & $4.33$  & $0.011$ & $2.53$  & $2.31\cdot10^{17}$ & $4.33$  & $0.011$ \\

L& 1.00   & -0.0 & $2.53$  & $2.31\cdot10^{17}$ & $4.33$  & $0.012$ & $2.53$  & $2.31\cdot10^{17}$ & $4.33$  & $0.012$ \\

L, S& 2.00   & -0.0 & $2.53$  & $2.31\cdot10^{17}$ & $4.33$  & $0.018$ & $2.53$  & $2.31\cdot10^{17}$ & $4.33$  & $0.018$ \\

S, C& 4.00   & -0.0 & $2.53$  & $2.31\cdot10^{17}$ & $4.33$  & $0.185$ & $2.53$  & $2.31\cdot10^{17}$ & $4.33$  & $0.185$ \\

C & 8.00   & -0.0 & $2.53$  & $2.31\cdot10^{17}$ & $4.33$  & $0.995$ & $2.53$  & $2.31\cdot10^{17}$ & $4.33$  & $0.995$ \\

~\\[-0.7em]

VHF&  0.10  & -5.0  &  $2.07$ & $7.43\cdot10^{15}$ & $0.78$  & $0.691$ & $2.21$ & $1.65\cdot10^{17}$ & $3.66$ & $0.082$ \\

UHF&  0.50  & -5.0   &  $2.07$ & $7.43\cdot10^{15}$ & $0.78$  & $0.622$ & $2.21$& $1.65\cdot10^{17}$ & $3.66$ & $0.038$ \\

L&  1.00  & -5.0   &  $2.07$ & $7.43\cdot10^{15}$ & $0.78$  & $0.925$ & $2.21$& $1.65\cdot10^{17}$ & $3.66$ & $0.040$ \\

L, S&  2.00  & -5.0   &  $2.07$ & $7.43\cdot10^{15}$ & $0.78$  & $0.985$ & $2.21$& $1.65\cdot10^{17}$ & $3.66$ & $0.060$ \\

S, C&  4.00  & -5.0  &  $2.07$ & $7.43\cdot10^{15}$ & $0.78$  & $0.996$ & $2.21$& $1.65\cdot10^{17}$ & $3.66$ & $0.899$ \\

C&  8.00  & -5.0  &  $2.07$ & $7.43\cdot10^{15}$ & $0.78$  & $0.999$ & $2.21$ & $1.65\cdot10^{17}$ & $3.66$ & $0.987$ \\

~\\[-0.7em]

VHF&  0.10  & -7.5 & $1.80$  &  $5.08\cdot10^{15}$ & $0.64$ & $0.642$  & $2.00$ & $1.14\cdot10^{17}$ & $3.04$ & $0.139$ \\

UHF & 0.50   & -7.5 & $1.80$  &  $5.08\cdot10^{15}$ & $0.64$ & $0.741$  & $2.00$ & $1.14\cdot10^{17}$ & $3.04$ & $0.085$ \\

L&  1.00  & -7.5 & $1.80$  &  $5.08\cdot10^{15}$ & $0.64$ & $0.990$  & $2.00$ &$1.14\cdot10^{17}$ & $3.04$ & $0.092$ \\

L, S&  2.00  & -7.5 & $1.80$  &  $5.08\cdot10^{15}$ & $0.64$ & $0.998$  & $2.00$ &$1.14\cdot10^{17}$ & $3.04$ & $0.149$ \\

S, C&  4.00  & -7.5 & $1.80$  &  $5.08\cdot10^{15}$ & $0.64$ & $0.999$  & $2.00$ & $1.14\cdot10^{17}$ & $3.04$ & $0.964$ \\

C & 8.00  & -7.5 & $1.80$  &  $5.08\cdot10^{15}$ & $0.64$ & $1.000$  & $2.00$ & $1.14\cdot10^{17}$ &$3.04$ & $0.994$ \\

\bottomrule
    \end{tabular*}
    \begin{tablenotes}
\item[{a}] The local plasma frequency in Hz is given by $f_{\rm p} = \frac{1}{2\pi}\sqrt{C_{\rm e}^2N_{\rm p}/(\varepsilon_0m_{\rm e})}$ or, after evaluating the constants, $f_{\rm p}\approx 9\sqrt{N_{\rm p}}$. 
\item[{b}] X--band communication (8-12 GHz) and above would be possible in all cases.

    \end{tablenotes}
   \end{threeparttable}
\end{table*}

\subsection{Effect of High Electric Field Correction to Ion Mobility}

We first examine the impact of incorporating the high electric field correction to the ion mobilities (see  Table~\ref{tab:mobilities:Ecorrection}). The application of the high electric field correction reduces the ion mobility within the cathode sheath by approximately one order of magnitude, and this leads to a two- or three-fold decrease in the sheath thickness, as can be observed in Fig.~\ref{fig:Ne_contours_Efield_effects_PM}. This is not surprising because we would theoretically expect such a significant decrease in ion mobility to necessitate a corresponding decrease in sheath thickness.

\begin{figure}[t]
    \centering
     \includegraphics[width=0.34\textwidth]{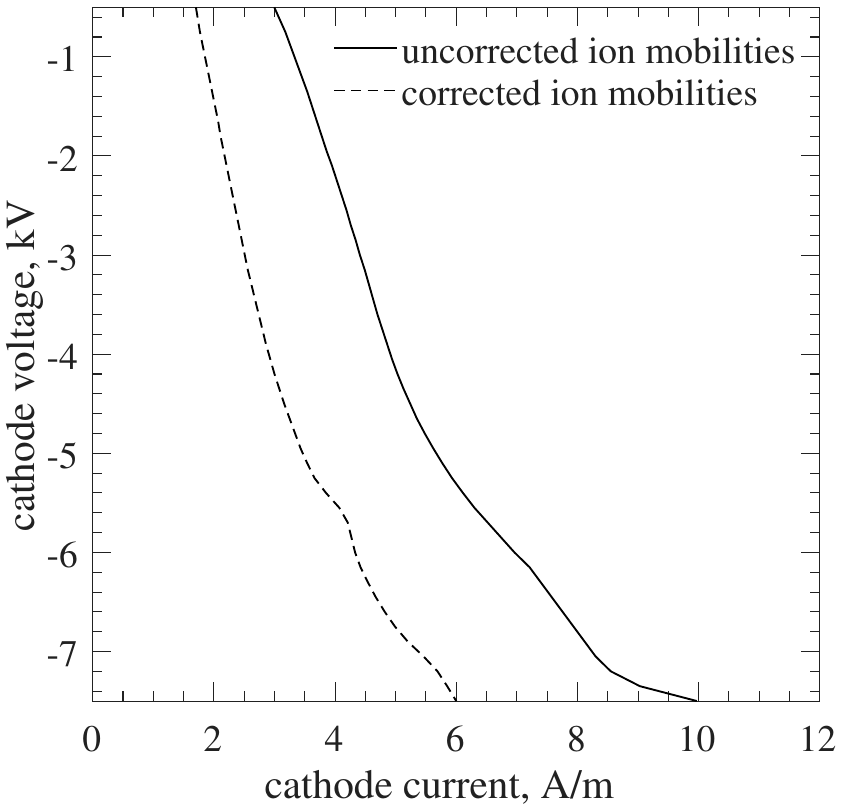}
     \figurecaption{Effect of ion mobility correction on the cathode voltage-current characteristic.}
     \label{fig:V_I_secA}
\end{figure}

This can be explained as follows. In a collisional cathode sheath characterized by a fixed voltage drop, the reduction of ion mobility leads to a contraction of the sheath thickness primarily due to the enhancement of space charge shielding. As the ion mobility decreases, the drift velocity of ions moving through the electric field is diminished, which necessitates a compensatory increase in the local ion number density to satisfy the continuity of current flux. This resulting accumulation of positive space charge increases the curvature of the electrostatic potential profile in accordance with Poisson's equation, thereby screening the cathode potential over a shorter distance. Consequently, the sheath compresses because the denser ion cloud is more efficient at shielding the electric field, a behavior consistent with the \cite{book:1948:mott} law (a.k.a.\ the collisional Child-Langmuir law) as outlined in \cite{book:2024:lieberman2024}:
\begin{equation}
J_{\rm i} \approx \frac{9}{8}\epsilon_0 \mu_{\rm i} \frac{\phi_{\rm s}^2}{d_{\rm s}^3}
\label{eqn:mottgurneylaw}
\end{equation}
where $J_{\rm i}$ is the ion current density, $\epsilon_0$ is the vacuum permittivity, $\phi_{\rm s}$ is the sheath voltage, and $d_{\rm s}$ is the cathode sheath thickness. Rearranging this relationship implies that the sheath thickness is proportional to the cubic root of the mobility ($d_{\rm s} \propto \mu_{\rm i}^{1/3}$). Therefore, should the ion mobility decrease by a factor of 10, the sheath thickness is expected to decrease by approximately a factor of 2. This contraction is clearly observed in the electron density contours presented in Figs.~\ref{fig:Ne_contours_Efield_effects_PM}b-c. It is important to note that the scaling $d_{\rm s} \propto \mu_{\rm i}^{1/3}$ assumes that the sheath voltage and ion current remain relatively constant despite changes in mobility. The potential contours in Fig.~\ref{fig:cathode_drop} confirm that the sheath voltage remains the same, as the voltage drop is largely confined to the sheath region regardless of the mobility model. Furthermore, while the current-voltage relationships in Fig.~\ref{fig:V_I_secA} indicate that the current decreases when mobility corrections are applied, the reduction is only by a factor of roughly 2. Although Fig.~\ref{fig:V_I_secA} presents the sum of the electronic and ionic currents, this value is effectively equal to the ionic current within the sheath, where the electron density is 4--5 orders of magnitude lower than the ion density. Consequently, this variation in current is secondary compared to the order-of-magnitude change in mobility, preserving the validity of the cubic-root scaling approximation.

These structural changes in the sheath due to a change in ion mobility model directly influence electromagnetic wave propagation. The estimates of $I/I_0$ are presented in Table~\ref{tab:atten}, comparing attenuation before the application of the pulsed electric field against the effects of the pulse using both uncorrected and corrected ion mobilities. Without the applied pulses, there is near-complete attenuation of L--band (1-2 GHz) and S--band (2-4 GHz) signals, with attenuation diminishing to 20\% or less for C--band (4-8 GHz) frequencies and above. The application of 7.5~kV pulses would theoretically permit L--band, S--band, and C--band communication with negligible attenuation ($I/I_0\approx 1$) if one assumes no high electric field effects on ion mobility. However, when the necessary high $E/N$ corrections are applied to the ion mobility, effective communication is restricted to the C--band and beyond.

It is emphasized that these attenuation results likely represent conservative lower bounds, as our physical model may underestimate the sheath thickness. Unlike a fully kinetic approach, the drift-diffusion model employed here assumes that ions remain in local equilibrium with the electric field. Even with high-field mobility corrections, the model forces ions to dissipate energy locally through collisions, capping their speed at the terminal drift velocity determined by the local $E/N$. This approximation effectively underestimates ion speeds in the bulk of the sheath. By ignoring the non-local nature of transport, the fluid model fails to account for the kinetic energy ions accumulate as they traverse multiple mean free paths across the potential drop. This results in an underestimated ion velocity, which requires a higher ion number density to satisfy flux conservation, leading to artificially strong shielding and a thinner sheath.

In a kinetic framework with discharge strengths exceeding 10,000 Td, ions do not fully thermalize with the background gas. While transport is not strictly ballistic, the mean free path in the heated sheath becomes a significant fraction of the sheath width. Consequently, ions accumulate directed energy from the macroscopic potential drop, attaining velocities significantly higher than the local collision-limited drift velocity. This enhanced velocity would result in a lower ion density for a given current, suppressed shielding, and a correspondingly thicker cathode sheath. We therefore expect that the sheath thickness predicted by the drift-diffusion model is somewhat underestimated due to the neglect of these non-local kinetic effects.

In addition to transport limitations, the use of the Local Field Approximation (LFA) for ionization may further contribute to the underestimation of sheath thickness. Because the Townsend ionization rates are functions of the local $E/N$, which is highest within the sheath, the LFA predicts significant ionization occurring inside the sheath region. This production source term introduces additional ions directly into the high-field region. The resulting enhancement of ion density further screens the applied electric field, artificially compressing the sheath.

\subsection{Effect of High Electric Field Correction to Electron Mobility}

We next isolate the influence of the electron mobility model on the sheath structure. In the baseline simulations, the electron mobility included corrections for high reduced electric fields ($E/N$). Here, we compare those results against a simplified model where the electron mobility $\mu_{\rm e}$ is strictly a function of the electron temperature $T_{\rm e}$, neglecting saturation effects due to high drift velocities. These time-accurate simulations were performed under the same driving conditions, utilizing a maximum pulse voltage of 7.5~kV.

The resulting communication window, shown in Fig.~\ref{fig:Ne_contours_Efield_effects_PM}d, exhibits negligible morphological differences compared to the baseline solution (Fig.~\ref{fig:Ne_contours_Efield_effects_PM}b). This insensitivity indicates that the electron mobility formulation has a minimal impact on the sheath thickness. Physically, this is a consequence of the cathode sheath being a highly non-neutral, electron-depleted region ($N_{\rm i} \gg N_{\rm e}$) due to the strong repulsive potential. The space charge density,  which dictates the curvature of the potential and the sheath width via Poisson's equation, is therefore determined almost exclusively by ion transport. Consequently, changes to the electron transport coefficients do not significantly alter the electrostatic shielding. This conclusion is consistent with the collisional Child-Langmuir scaling discussed in Eq.~(\ref{eqn:mottgurneylaw}), which explicitly defines the sheath thickness $d_{\rm s}$ as a function of the ion mobility $\mu_{\rm i}$, independent of $\mu_{\rm e}$.

\begin{figure}[!b]
     \subfigure[$\gamma_{\rm e}=0.1$]{\includegraphics[width=0.49\textwidth]{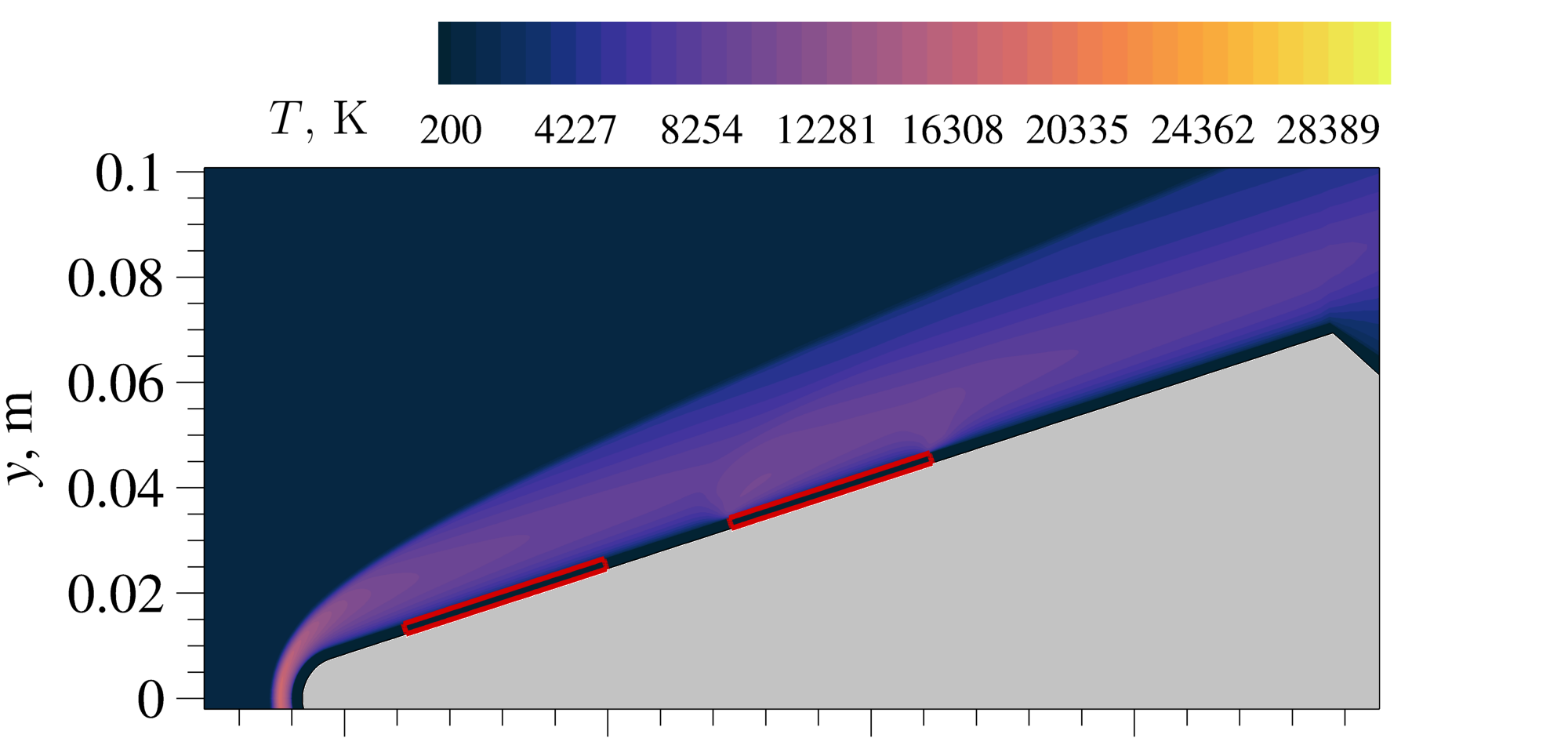}}
     \subfigure[$\gamma_{\rm e}=0.5$]{\includegraphics[width=0.49\textwidth]{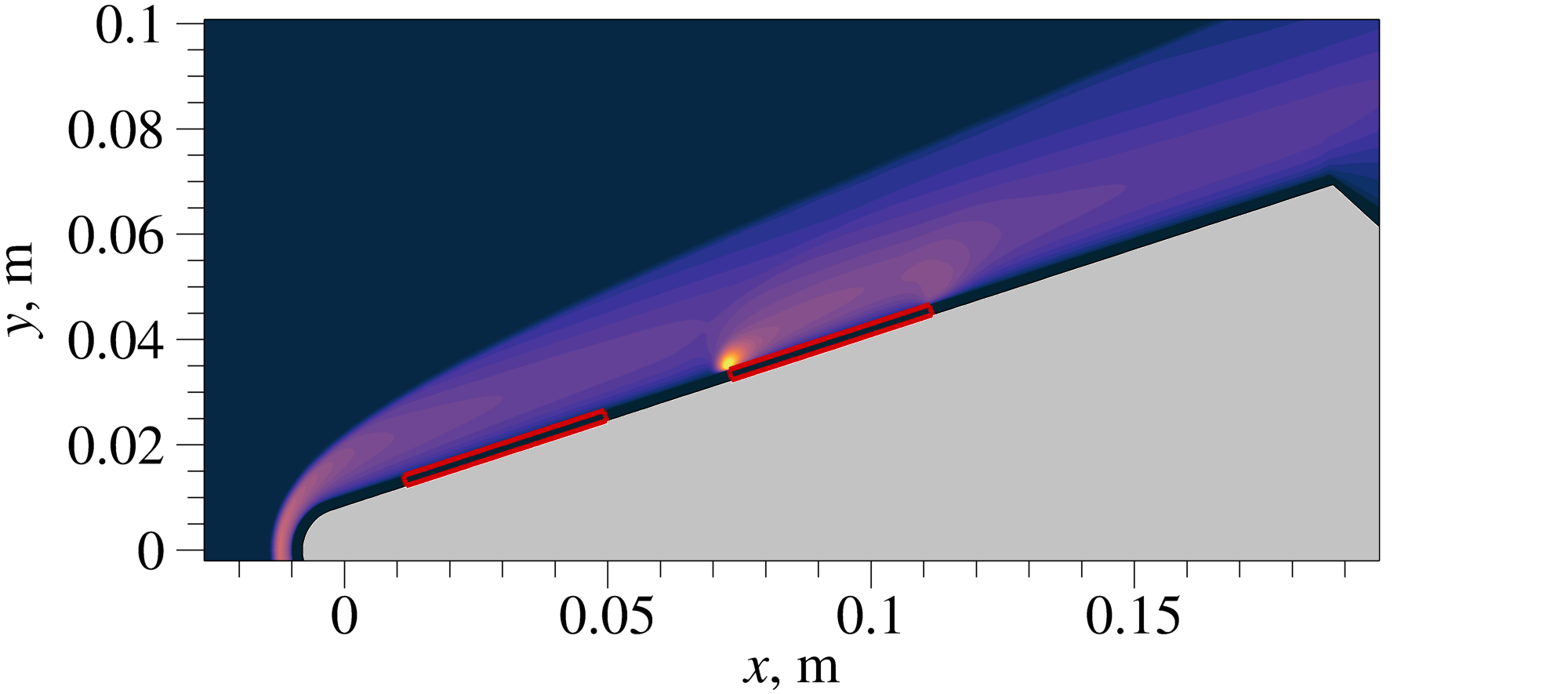}}
     \figurecaption{Bulk plasma temperature contours at peak voltage (7.5~kV) for secondary electron emission coefficients of (a) $\gamma_e = 0.1$ and (b) $\gamma_e = 0.5$.}
     \label{fig:thermal_runaway_gammae05}
\end{figure}

\subsection{Effect of Secondary Electron Emission}
\begin{figure}[!t]
     \centering
     \subfigure[Electron number density]{\includegraphics[width=0.37\textwidth]{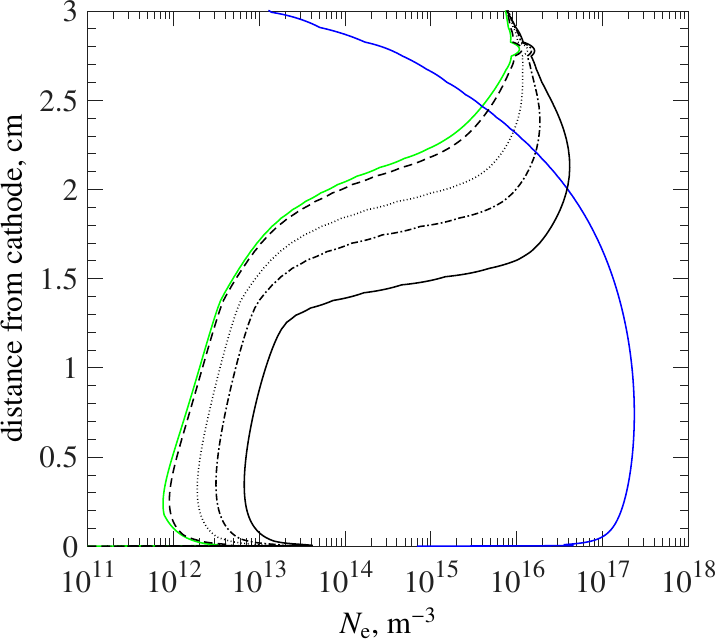}}
     \subfigure[Power deposited]{\includegraphics[width=0.37\textwidth]{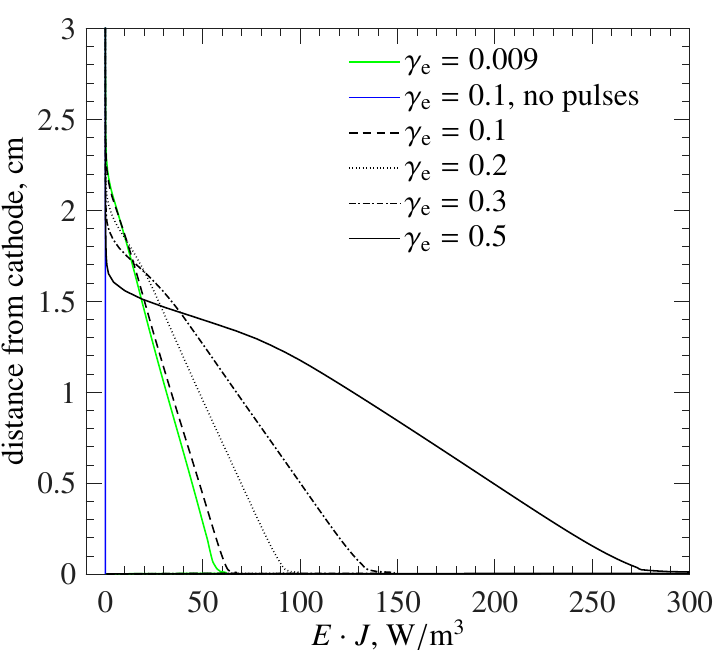}}
     \subfigure[Gas temperature]{\includegraphics[width=0.38\textwidth]{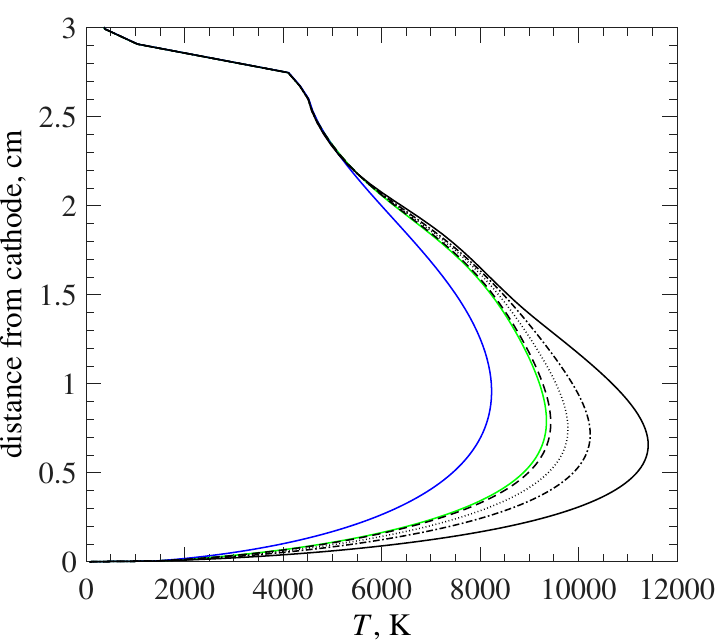}}
     \figurecaption{Effect of secondary emission coefficient $\gamma_{\rm e}$ on discharge properties at peak voltage at midpoint of cathode:  (a) Electron density; (b) Power deposited; (c) Gas temperature. The properties are extracted from the midpoint of the cathode, at the instant where the applied voltage is 7.5~kV.}
     \label{fig:SEE_effects}
\end{figure}

Here, we investigate the influence of the secondary electron emission (SEE) coefficient, $\gamma_{\rm e}$, on the sheath structure and flowfield properties. {The determination of $\gamma_{\rm e}$ is subject to significant uncertainty, as it depends heavily on the cathode material, surface cleanliness, reduced electric field ($E/N$), and the flux of metastables. As noted by \cite{psst:1999:phelps}, $\gamma_{\rm e}$ can vary by orders of magnitude (from 0.01 to 5 in argon) depending on these surface conditions. For nitrogen discharges, values around 0.1 are commonly cited in literature; however, our validation against the experimental data of \cite{thesis:2022:broslawski} (see Section IV-C) required a value of $\gamma_{\rm e}=0.5$ to match the voltage-current characteristics in a high-$E/N$ regime. We note that while the validation case shares similar gas densities and electric fields with the present study, significant differences exist in the wall temperature (approx.~300--450~K in experiments vs. $>1000$~K in re-entry) and the population of vibrationally excited species. Given these uncertainties,} we vary $\gamma_{\rm e}$ over a representative range of 0.1 to 0.5 {to bracket the potential outcomes}.

A primary consequence of increasing the SEE coefficient is the enhancement of Joule heating, which drives higher gas temperatures. Physically, a higher $\gamma_{\rm e}$ increases the flux of electrons emitted from the cathode into the sheath. This effectively lowers the impedance of the sheath region, allowing for a larger total discharge current for a fixed voltage. The resulting increase in power deposition ($\vec{E} \cdot \vec{J}$) exacerbates gas heating. {This heating is particularly intense at the electrode edges due to the electrostatic concentration of electric field lines (the edge effect). The resulting local enhancement of the electric field drives a spike in current density,} which exacerbates gas heating. This effect is evident in Figs.~\ref{fig:thermal_runaway_gammae05}, \ref{fig:SEE_effects}b, and \ref{fig:SEE_effects}c, where increasing $\gamma_{\rm e}$ from 0.1 to 0.5 elevates gas temperatures at the electrode edge beyond 20,000 K.

The sheath thickness also responds to variations in the SEE coefficient, though the sensitivity is less acute than that of the gas temperature. As illustrated in Fig.~\ref{fig:SEE_effects}a, a fivefold increase in $\gamma_{\rm e}$ results in a sheath contraction of approximately 30\%. Theoretical considerations suggest that the macroscopic sheath thickness should remain relatively robust against changes in electron emission. This robustness stems from the extreme velocity disparity between the massive ions and the accelerated electrons; because the local space charge density scales inversely with velocity, the rapidly escaping secondary electrons provide negligible direct screening compared to the slow, dense ion background. This holds true provided the emission remains below the critical space-charge limited threshold (typically $\gamma_{\rm e} \approx 1$), at which point the surface electric field would collapse.

The moderate reduction in sheath thickness observed in our simulations is therefore likely exaggerated by the Local Field Approximation (LFA) inherent to the fluid model. Under the LFA, the ionization source term is strictly a function of the local reduced electric field ($E/N$). Since $E/N$ peaks within the high-voltage sheath, the model predicts an intense, localized ionization avalanche within the sheath itself. An increase in $\gamma_{\rm e}$ injects more seed electrons into this high-multiplication zone, creating a denser population of locally generated ions adjacent to the cathode. These ions enhance the screening of the electrode potential, artificially compressing the sheath.

In a fully kinetic or non-local physical framework, we expect the sheath thickness to be far less sensitive to the SEE coefficient. At the high reduced electric fields characteristic of the cathode fall (often exceeding $10^4$ Td), emitted electrons rapidly enter the runaway regime where the ionization cross-section decreases with increasing energy. These electrons traverse the sheath ballistically, depositing the majority of their energy in the negative glow rather than within the sheath itself. Because the sheath remains essentially a source-free `dark' region in this regime, the ion density profile is determined primarily by the flux entering from the bulk plasma rather than by local production. Consequently, the sheath thickness should remain largely independent of $\gamma_{\rm e}$ until the emitted electron flux becomes comparable to the ion flux.

\subsection{Feasibility and Comparative Assessment}

{
The simulations presented herein predict a power density requirement of approximately 66 W/cm$^2$ to sustain the electron depletion layer. While this flux appears significant, the total system load must be evaluated in the context of the limited surface area and duration required for telemetry. For a practical re-entry vehicle, a cathode surface area of 50~cm$^2$ is sufficient to establish a transmission window with a broad communication angle. This results in a total instantaneous power requirement of approximately 3.3~kW.}

{Considering that the radio blackout phase typically lasts between 2 and 10 minutes (depending on the vehicle geometry and trajectory), the total energy required to maintain the window ranges from 110~Wh to 550~Wh. With modern aerospace-grade lithium-ion battery packs offering specific energies exceeding 200~Wh/kg, the dedicated power supply for this system would weigh between 0.5~kg and 3.0~kg. Furthermore, mass and volume requirements can be reduced by an order of magnitude if the communication window is opened intermittently. By buffering telemetry data and transmitting in bursts (e.g., activating the depletion field for 1 second every 10 seconds), the average power consumption drops significantly, rendering the battery mass truly negligible (on the order of hundreds of grams).}

{This modest weight penalty presents a significant advantage over alternative blackout mitigation strategies. For instance, magnetic window concepts typically require heavy superconducting magnets or electromagnets with prohibitive power consumption to penetrate the high-density plasma layer. Similarly, electrophilic liquid injection systems incur substantial weight and complexity penalties associated with fluid reservoirs, plumbing, and active flow control. In contrast, the proposed pulsed electrostatic method offers a lightweight, solid-state solution that is mechanically simple and feasible for implementation on standard re-entry capsules.}

\section{Conclusions}

{Using an established and advanced numerical framework,} this study presented the first fully-coupled, reacting flow simulations investigating the use of high-voltage pulsed electric fields to locally deplete electron density in hypersonic re-entry flows. By applying $7.5$ kV pulses to electrodes on a Mach 24 waverider at an altitude of 68 km, we demonstrated the formation of a substantial non-neutral plasma sheath that effectively acts as a communication window. The simulations indicate that this technique can reduce the attenuation of a 4 GHz signal from a prohibitive 60\% to a transmissive 4\%, requiring a manageable power per surface area of approximately $66\text{ W/cm}^{2}$. {Our analysis confirms that the required energy storage for a typical blackout phase corresponds to a battery mass of 0.5 to 3.0 kg, or significantly less if intermittent transmission strategies are employed.} These results validate the feasibility of electrostatic mitigation strategies within a fully coupled aerodynamic and plasma discharge framework.

A rigorous sensitivity analysis of the transport and surface models revealed that the sheath topology is governed primarily by ion kinetics rather than electron transport. We found that incorporating high electric field corrections to ion mobility is essential for accurate modeling, as the reduced mobility in the high-field sheath leads to increased space charge shielding and a subsequent contraction of the sheath thickness. Conversely, the sheath structure proved remarkably insensitive to the specific electron mobility model employed, confirming that the potential drop is sustained almost exclusively by the ion space charge. Furthermore, variations in the secondary electron emission (SEE) coefficient highlighted a thermal constraint; while the sheath thickness is only moderately affected by the SEE coefficient in this fluid approximation, higher emission yields significantly enhanced Joule heating, potentially leading to excessive temperatures and heat fluxes at the cathode leading edge.

Finally, our analysis identifies three distinct physical mechanisms in the present drift-diffusion model that likely lead to an underestimation of the sheath thickness. First, the fluid assumption of local equilibrium ignores ion inertia, preventing ions from attaining directed velocities significantly higher than the collision-limited drift velocity. Second, the Local Field Approximation over predicts ionization within the high-field sheath region, artificially pumping too many ions into the sheath. Third, secondary electrons in a fluid model contribute to this local ionization avalanche, whereas in a kinetic framework, they would traverse the sheath ballistically and deposit their energy in the negative glow. All three limitations of the fluid approach act constructively to overestimate ion density and underestimate sheath thickness in our results. Consequently, a higher-fidelity kinetic approach is expected to yield significantly thicker sheaths for equivalent power deposition. This implies that the actual mitigation performance may well exceed the results presented here, yielding even lower signal attenuation and further reinforcing the viability of using pulsed electric fields to maintain communication during hypersonic re-entry.

\section*{Data Availability}

The data that support the findings of this study are available from
the corresponding author upon reasonable request.
\footnotesize
\bibliography{all}

@conference{nasa:1990:gupta,
        author = "Gupta, R. N. AND Yos, J. M. AND Thompson, R. A. AND Lui, K. P.",
        TITLE = "A Review of Reaction
Rates and Thermodynamic and Transport Properties for an 11-Species Air Model for Chemical and Thermal Nonequilibrium
Calculations to 30 000 K",
        booktitle = "NASA RP-1232",
        year = 1990,
        month = "Aug"
        }

@conference{aiaaconf:1987:bardina,
        author = "Bardina, J. E. AND Lombard, C. K.",
        title = "Three Dimensional Hypersonic Flow
       Simulations with the CSCM Implicit Upwind Navier-Stokes Method",
        booktitle = "Proceedings of the 8th AIAA Computational Fluid Dynamics Conference" ,
        note = "AIAA Paper 87-1114" ,
        year = 1987,
        doi = {10.2514/6.1987-1114}
        }

@article{jcp:2014:parent,
        author = "Parent, B. AND Macheret, S. O. AND Shneider, M. N. ",
        TITLE = "Electron and Ion Transport Equations in Computational Weakly-Ionized Plasmadynamics",
        journal = "Journal of Computational Physics",
		  year = 2014,
        volume = 259,
        pages = "51--69",
        doi = {10.1016/j.jcp.2013.11.029}
        }

@article{aiaa:2016:parent,
        author = "Parent, B. AND Shneider, M. N. AND Macheret, S. O.",
        title = "Detailed Modeling of Plasmas for Computational Aerodynamics",
        journal = "AIAA Journal",
        year = 2016,
        volume = 54,
        pages = "898--911",
        number = 3,
        doi = {10.2514/1.J054624}
        }

@book{book:1991:raizer,
        author = "Raizer, Yu. P.",
        title = "Gas Discharge Physics",
        publisher = "Springer-Verlag",
	address = "Berlin, Germany",
        year = 1991,
doi={10.1007/978-3-642-61247-3}
        }

@article{jsr:2008:keidar,
  title={Electromagnetic reduction of plasma density during atmospheric reentry and hypersonic flights},
  author={Keidar, Michael and Kim, Minkwan and Boyd, Iain D},
  journal={Journal of Spacecraft and Rockets},
  volume={45},
  number={3},
  pages={445--453},
  year={2008},
  doi={10.2514/1.32147}
}

@article{pf:2007:boyd,
	Author = {Boyd,Iain D.},
	Doi = {10.1063/1.2771662},
	Journal = {Physics of Fluids},
	Number = {9},
	Pages = {096102},
	Title = {Modeling of associative ionization reactions in hypersonic rarefied flows},
	Volume = {19},
	Year = {2007}
}

@article{jcp:1996:bose,
	Author = {Bose,Deepak and Candler,Graham V.},
	Doi = {10.1063/1.471106},
	Journal = {The Journal of Chemical Physics},
	Number = {8},
	Pages = {2825-2833},
	Title = {Thermal rate constants of the $\rm N_2+O\rightarrow NO + N$ reaction using ab initio 3A and 3A potential energy surfaces},
	Volume = {104},
	Year = {1996}
}

@article{jcp:1997:bose,
	Author = {Bose,Deepak and Candler,Graham V.},
	Doi = {10.1063/1.475132},
	Journal = {The Journal of Chemical Physics},
	Number = {16},
	Pages = {6136-6145},
	Title = {Thermal rate constants of the $\rm O_2+N\rightarrow NO + O$ reaction based on the A2 and A4 potential-energy surfaces},
	Volume = {107},
	Year = {1997}
}

@book{book:1990:park,
        author = "Park, C.",
        title = "Nonequilibrium Hypersonic Aerothermodynamics",
        publisher = "Wiley, New-York",
        year = 1990,
        doi={10.1063/1.2809999}
        }

@book{book:1997:grigoriev,
        author = "Grigoriev, I. S. AND Meilikhov, E. Z.",
        title = "Handbook of Physical Quantities",
        publisher = "CRC",
	address = "Boca Raton, Florida",
        year = 1997
        }

@article{aiaa:1998:roache,
        author = "Roache, P. J.",
        title = "Verification of Codes and Calculations",
journal = {AIAA Journal},
volume = {36},
number = {5},
pages = {696-702},
year = {1998},
doi = {10.2514/2.457},
        }

@article{cf:2001:maccormack,
        author = "MacCormack, R. W.",
        TITLE = "Iterative Modified Approximate Factorization",
        journal = "Computers and Fluids",
        volume = 30,
        number=8,
        pages = "917--925",
        year = 2001,
        doi = {10.1016/S0045-7930(01)00035-4}
        }

@article{gen:douglas,
        author = "Douglas, Jr., J.",
        TITLE = "On the Numerical Integration of $\partial^2 u / \partial x^2
                    +\partial^2 u / \partial y^2=\partial u / \partial t$
                    by Implicit Methods",
        journal = "J. Soc. Ind. Appl. Math." ,
        year = 1955,
        volume = 3,
        pages = "42--65",
        doi={10.1137/0103004}
        }

@article{misc:1968:sinnott,
        author = "Sinnott, G. AND Golden, D. E. AND Varney, R. N.",
        TITLE = "Positive-Ion Mobilities in Dry Air",
        journal = "Physical Review",
        volume = 170,
        number = 1,
        pages = "272--275",
        year = 1968,
        doi = {10.1103/PhysRev.170.272}
        }

@article{jcp:2015:parent,
  title={Modeling weakly-ionized plasmas in magnetic field: A new computationally-efficient approach},
  author={Parent, Bernard and Macheret, Sergey O and Shneider, Mikhail N},
  journal={Journal of Computational Physics},
  volume={300},
  pages={779--799},
  year={2015},
  publisher={Elsevier},
  doi = {10.1016/j.jcp.2015.08.010}
}

@conference{nasa:1970:grantham,
        author = "Grantham, W. L.",
        title = "Flight Results of a 25 000-Foot-Per-Second Reentry Experiment Using Microwave Reflectometers To Measure Plasma Electron Density and Standoff Distance",
        booktitle = "TN D-6062, NASA",
        year = 1970
        }

@conference{nasa:1972:jones,
        author = "Jones, W. and Cross, A.",
        title = "Electrostatic-probe measurements of plasma parameters for two reentry flight experiments at 25000 feet per second",
        booktitle = "TN D-6617, NASA",
        year = 1972,
        month = 03
}

@incollection{book:2022:parent,
        author = "Parent, B.",
        editor = "Colonna, G.  and D'Angola, A.",
        edition= "2nd",
        title = "Drift-Diffusion Models and Methods",
        booktitle = "Plasma Modeling: Methods and Applications",
        chapter = 8,
        publisher = "IOP Publishing",
        year = 2022,
        doi = "10.1088/978-0-7503-3559-1ch8",
        }

@conference{nasa:2002:mcbride,
        author = "McBride, B. J. AND Zehe, M. J. AND Gordon, S.",
        title = "NASA Glenn Coefficients for Calculating Thermodynamic Properties of Individual Species",
        booktitle = "NASA TP--2002-211556",
        year = 2002,
        month = "Sep"
        }

@conference{nasa:1973:dunn,
        author = "Dunn, M. G. AND Kang, S. W.",
        TITLE = "Theroretical and Experimental Studies of Reentry Plasmas",
        booktitle = "NASA CR-2232",
        year = 1973,
        }

@article{jtht:2023:parent,
  title={Effect of Plasma Sheaths on Earth-Entry Magnetohydrodynamics},
  author={Parent, Bernard and Thoguluva Rajendran, Prasanna and Macheret, Sergey O and Little, Justin and Moses, Robert W and Johnston, Christopher O and Cheatwood, F McNeil},
  journal={Journal of thermophysics and heat transfer},
  volume={37},
  number={4},
  pages={845--857},
  year={2023},
  publisher={American Institute of Aeronautics and Astronautics},
  doi={10.2514/1.T6784}
}

@inproceedings{asvpaper:1995:inouye,
  title={OREX flight--Quick report and lessons learned},
  author={Inouye, Yasutoshi},
  booktitle={Aerothermodynamics for space vehicles, Proceedings of the 2nd European Symposium, 21-25 November, Published by European Space Agency},
  volume={367},
  pages={271},
  year={1995}
}

@article{sw:2002:doihara,
        author = {R. Doihara and M. Nishida},
        title = "Thermochemical nonequilibrium viscous shock layer studies of the orbital reentry experiment ({OREX}) vehicle",
        journal = "Shock Waves",
	    year = 2002,
	    volume = 11,
        number = 5,
        pages = "331--339"
        }

@article{jtht:1993:park,
        author = "Chul Park",
        title = "Review of chemical-kinetic problems of future {NASA} missions. I - Earth entries",
    	journal = "Journal of Thermophysics and Heat Transfer",
    	year = 1993,
    	month = {jul},
    	volume = 7,
    	number = 3,
    	pages = "385--398",
        doi={10.2514/3.431}    
        }

@techreport{nrl:2002:huba,
        author = "Huba, J.D.",
        TITLE = "NRL Plasma Formulary",
        institution = "NRL",
        year = 2002
        }

@article{jgr:2004:sheehan,
  title={Dissociative recombination of $\rm N_2^+$, $\rm O_2^+$, and $\rm NO^+$: Rate coefficients for ground state and vibrationally excited ions},
  author={Sheehan, Clint H and St.-Maurice, J-P},
  journal={Journal of Geophysical Research: Space Physics},
  volume={109},
  number={A3},
  year={2004},
  publisher={Wiley Online Library},
  doi={10.1029/2003JA010132} 
}

@article{aip:1998:peterson,
  title={Dissociative recombination and excitation of $\rm N_2^+$: Cross sections and product branching ratios},
  author={Peterson, JR and Le Padellec, A and Danared, H and Dunn, GH and Larsson, M and Larson, A and Peverall, R and Str{\"o}mholm, C and Ros{\'e}n, S and Af Ugglas, M and others},
  journal={Journal of Chemical Physics},
  volume={108},
  number={5},
  pages={1978--1988},
  year={1998},
  publisher={American Institute of Physics},
  doi={10.1063/1.475577}
}

@article{aip:2001:peverall,
  title={Dissociative recombination and excitation of $\rm O_2^+$: Cross sections, product yields and implications for studies of ionospheric airglows},
  author={Peverall, Robert and Ros{\'e}n, Stefan and Peterson, James R and Larsson, Mats and Al-Khalili, Ahmed and Vikor, Ljiljana and Semaniak, Jacek and Bobbenkamp, Rolf and Le Padellec, Arnaud and Maurellis, AN and others},
  journal={Journal of Chemical Physics},
  volume={114},
  number={15},
  pages={6679--6689},
  year={2001},
  publisher={American Institute of Physics},
  doi={10.1063/1.1349079} 
}

@article{pf:2024:parent,
  title={Progress in electron energy modeling for plasma flows and discharges},
  author={Parent, Bernard and Rodríguez Fuentes, Felipe Martin},
  journal={Physics of Fluids},
  volume={36},
  number={8},
  year={2024},
  publisher={AIP Publishing},
 doi = {10.1063/5.0219552}
}

@article{jpcr:1991:phelps,
  title={Cross sections and swarm coefficients for nitrogen ions and neutrals in $\rm N_2$ and argon ions and neutrals in Ar for energies from 0.1 eV to 10 keV},
  author={Phelps, Av V},
  journal={Journal of Physical and Chemical Reference Data},
  volume={20},
  number={3},
  pages={557--573},
  year={1991},
doi = {10.1063/1.555889}
}

@inproceedings{pr:1954:wannier,
  title={Corrected values for the charge transfer cross section in the noble gases},
  author={Wannier, GH},
  booktitle={Physical Review},
  volume={96},
  number={3},
  pages={831--831},
  year={1954},
  organization={American Physical Society}
}

@article{jap:2023:pokharel,
  title={Self-consistent model and numerical approach for laser-induced non-equilibrium plasma},
  author={Pokharel, S and Tropina, AA},
  journal={Journal of Applied Physics},
  volume={134},
  number={22},
  year={2023},
  publisher={AIP Publishing},
  doi = {10.1063/5.0175177}
}

@phdthesis{thesis:2022:broslawski,
  title={The modeling and experimentation of hypersonic turbulent boundary layers with and without thermal nonequilibrium},
  author={Broslawski, Casey Joseph},
  year={2022},
  school={Texas A\&M University}
}

@article{jap:2019:peters,
  title={Kinetics model of femtosecond laser ionization in nitrogen and comparison to experiment},
  author={Peters, Christopher J and Shneider, Mikhail N and Miles, Richard B},
  journal={Journal of applied physics},
  volume={125},
  number={24},
  year={2019},
  publisher={AIP Publishing},
doi = {10.1063/1.5098306}
}

@book{book:1962:barrow,
title = {Introduction to Molecular Spectroscopy},
author = {G. M. Barrow},
publisher = {Acta Crystallographica},
volume = {16},
year = {1963},
doi = {10.1107/S0365110X63003467},
}

@article{jap:2007:korotkevich,
  title={Communication through plasma sheaths},
  author={Korotkevich, AO and Newell, AC and Zakharov, VE},
  journal={Journal of Applied Physics},
  volume={102},
  number={8},
  year={2007},
  publisher={AIP Publishing},
  doi={10.1063/1.2794856}
}

@inproceedings{ieee:2017:steffens,
  title={Experimental verification of pulsed electrostatic manipulation for reentry blackout alleviation},
  author={Steffens, Lars and Krishnamoorthy, Siddharth and G{\"u}lhan, Ali and Close, Sigrid},
  booktitle={2017 IEEE Aerospace Conference},
  pages={1--12},
  year={2017},
  organization={IEEE},
  doi={10.1109/AERO.2017.7943891}
}

@article{psst:2024:luo,
  title={Ground experimental study of the electron density of plasma sheath reduced by pulsed discharge},
  author={Luo, Cheng and Zhang, Jia and Liu, Yanming and Wei, Qiang and Dang, Mengjia and Ba, Yongshan and Gao, Jingru and Li, Yuxin},
  journal={Plasma Sources Science and Technology},
  volume={33},
  number={9},
  year={2024},
  publisher={IOP Publishing},
  doi={10.1088/1361-6595/ad75b3}
}

@article{jsr:2010:kim,
  title={Modeling of electromagnetic manipulation of plasmas for communication during reentry flight},
  author={Kim, Minkwan and Boyd, Iain D and Keidar, Michael},
  journal={Journal of Spacecraft and Rockets},
  volume={47},
  number={1},
  pages={29--35},
  year={2010},
  doi={10.2514/1.45525}
}

@article{pf:2025:rodriguez2,
    author = {Rodr\'{i}guez Fuentes, Felipe Martin and Parent, Bernard},
    title = {Vibrational-electron heating in plasma flows: A thermodynamically consistent model},
    journal = {Physics of Fluids},
    volume = {37},
    number = {9},
    pages = {096141},
    year = {2025},
    doi = {10.1063/5.0285170}
}

@article{ps:2005:tarasenko,
  title={High-power subnanosecond beams of runaway electrons generated in dense gases},
  author={Tarasenko, Victor F and Yakovlenko, Sergei I},
  journal={Physica scripta},
  volume={72},
  number={1},
  pages={41},
  year={2005},
  publisher={IOP Publishing},
  doi={10.1238/Physica.Regular.072a00041}
}

@article{sw:1994:toro,
  title={Restoration of the contact surface in the HLL-Riemann solver},
  author={Toro, Eleuterio F and Spruce, Michael and Speares, William},
  journal={Shock waves},
  volume={4},
  number={1},
  pages={25--34},
  year={1994},
  publisher={Springer},
  doi={10.1007/BF01414629}
}

@article{siam:1983:harten,
  title={On upstream differencing and Godunov-type schemes for hyperbolic conservation laws},
  author={Harten, Amiram and Lax, Peter D and Leer, Bram van},
  journal={SIAM review},
  volume={25},
  number={1},
  pages={35--61},
  year={1983},
  publisher={SIAM},
  doi = {10.1137/1025002}
}

@inproceedings{aiaapaper:1996:gupta,
  title={Assessment of thermochemical nonequilibrium and slip effects for Orbital Reentry Experiment (OREX)},
  author={Gupta, Roop N. and Moss, James N. and Price, Joseph M.},
  booktitle={31st Thermophysics Conference},
  year={1996},
  month={jun},
  publisher={American Institute of Aeronautics and Astronautics},
  address={New Orleans, LA},
  doi={10.2514/6.1996-1859},
  note={AIAA Paper 96-1859}
}

@article{book:1948:mott,
  title={Electronic processes in ionic crystals},
  author={Mott, Nevill Francis and Gurney, Ronald Wilfrid},
  year={1948}
}

@book{book:2024:lieberman2024,
  title={Principles of Plasma Discharges and Materials Processing},
  author={Lieberman, Michael A and Lichtenberg, Allan J},
  year={2024},
  publisher={John Wiley \& Sons}
}

@article{cf:2018:simon,
  title={A cure for numerical shock instability in HLLC Riemann solver using antidiffusion control},
  author={Simon, Sangeeth and Mandal, JC},
  journal={Computers \& Fluids},
  volume={174},
  pages={144--166},
  year={2018},
  publisher={Elsevier},
  doi={10.1016/j.compfluid.2018.07.001}
}

@article{nasa::1994:gillman,
  title={Review of leading approaches for mitigating hypersonic vehicle communications blackout and a method of ceramic particulate injection via cathode spot arcs for blackout mitigation},
  author={Gillman, Eric D and Foster, John E and Blankson, Isaiah M},
  journal = "NASA TM-2010-216220" ,
  year = 2010
        }

@article{nasa::1968:schroeder,
  title={Flight investigation and analysis of alleviation of communications blackout by water injection during Gemini 3 reentry},
  author={Schroeder, Lyle C and Russo, Francis P},
  journal={NASA TM X-1521},
  pages={1--56},
  year={1968}
        }

@inproceedings{aiaaconf:2023:sawicki,
  title={Effect of water vapor injection on plasma reduction in hypersonic flow},
  author={Sawicki, Pawel and Campbell, Nicholas S and Boyd, Iain D},
  booktitle={AIAA SCITECH 2023 Forum},
  year={2023}
}

@inproceedings{rast:2011:kim,
  title={Plasma manipulation using a MHD-based device for a communication blackout in hypersonic flights},
  author={Kim, Minkwan and G{\"u}lhan, Ali},
  booktitle={Proceedings of 5th International Conference on Recent Advances in Space Technologies-RAST2011},
  pages={412--417},
  year={2011},
  organization={IEEE},
  doi={10.1109/RAST.2011.5966868}
}

@article{jpd:2017:krishnamoorthy,
  title={Investigation of plasma--surface interaction effects on pulsed electrostatic manipulation for reentry blackout alleviation},
  author={Krishnamoorthy, S and Close, S},
  journal={Journal of Physics D: Applied Physics},
  volume={50},
  number={10},
  pages={105202},
  year={2017},
  publisher={IOP Publishing},
  doi={10.1088/1361-6463/aa5901}
}

@inproceedings{aiaaconf:2025:anderson,
  title={Analysis of Magnetic Field Effects on Radio Signal Propagation Through Hypersonic Layer Plasma},
  author={Anderson-Ciccone, Eben and Riley, Landon and Maikov, Maksim and Tropina, Albina and Miles, Richard B},
  booktitle={AIAA AVIATION FORUM AND ASCEND 2025},
  year={2025},
  doi={10.2514/6.2025-3849}
}

@article{ijms:2009:krylov,
  title={Electric field dependence of the ion mobility},
  author={Krylov, EV and Nazarov, EG},
  journal={International Journal of Mass Spectrometry},
  volume={285},
  number={3},
  pages={149--156},
  year={2009},
  publisher={Elsevier},
  doi={10.1016/j.ijms.2009.05.009}
}

@article{kp:1987:mnatsakanyan,
  title={Processes of formation and decay of charged particles in nitrogen-oxygen plasmas},
  author={Mnatsakanyan, A Kh and Naidis, GV},
  journal={Khimiia Plazmy [Plasma Chemistry]},
  volume={14},
  pages={227--255},
  year={1987}
}

@article{zfp:1965:schlumbohm,
  title={Sto{\ss}ionisierungskoeffizient $\alpha$, mittlere Elektronenenergien und die Beweglichkeit von Elektronen in Gasen},
  author={Schlumbohm, Hans},
  journal={Zeitschrift f{\"u}r Physik},
  volume={184},
  number={5},
  pages={492--505},
  year={1965},
  publisher={Springer},
  doi = {10.1007/BF01380592}
}

@article{atomicdata:1976:ellis,
  title={Transport properties of gaseous ions over a wide energy range},
  author={Ellis, HW and Pai, RY and McDaniel, EW and Mason, EA and Viehland, LA},
  journal={Atomic data and nuclear data tables},
  volume={17},
  number={3},
  pages={177--210},
  year={1976},
  publisher={Elsevier},
 doi={10.1016/0092-640X(76)90001-2}
}

@techreport{nasa:1989:gnoffo,
  title={Conservation equations and physical models for hypersonic air flows in thermal and chemical nonequilibrium},
  author={Gnoffo, Peter A},
  volume={2867},
  year={1989},
  publisher={National Aeronautics and Space Administration, Office of Management.}
}

@article{cpc:2006:jin,
  title={Effects of external magnetic field on propagation of electromagnetic wave in uniform magnetized plasma slabs},
  author={Jin, Fanya and Tong, Honghui and Shi, Zhongbing and Tang, Deli and Chu, Paul K},
  journal={Computer physics communications},
  volume={175},
  number={8},
  pages={545--552},
  year={2006},
  publisher={Elsevier},
 doi={10.1016/j.cpc.2006.06.010}
}

@article{psst:1999:phelps,
  title={Cold-cathode discharges and breakdown in argon: surface and gas phase production of secondary electrons},
  author={Phelps, AV and Petrovic, Z Lj},
  journal={Plasma Sources Science and Technology},
  volume={8},
  number={3},
  pages={R21},
  year={1999},
  publisher={IOP Publishing},
 doi = {10.1088/0963-0252/8/3/201}
}

@article{htmp:2005:enzian,
  title={Post-flight analyses of the OREX catalycity experiment},
  author={Enzian, A and Ito, T and Balat-Pichelin, M and Desportes, A and Vervisch, P and Guyon, C and Amouroux, Jacques and Tran, Ph and Sauvage, N and Thivet, F},
  journal={High Temperature Material Processes: An International Quarterly of High-Technology Plasma Processes},
  volume={9},
  number={3},
  year={2005},
  publisher={Begel House Inc.},
  doi={10.1615/HighTempMatProc.v9.i3.30}
}
\bibliographystyle{plainnatmod}
\end{document}